\documentclass[
]{aa}  

\usepackage{graphicx,natbib}
\usepackage{txfonts}
\usepackage{dcolumn}
\usepackage{lscape}
\usepackage{amsmath}
\usepackage{booktabs}
\usepackage{color, colortbl}
\definecolor{ratgray}{gray}{0.9}
\definecolor{dblue}{rgb}{0.0, 0.0, 0.65}
\usepackage[
    pdfmenubar=true,
    colorlinks=true,
    linkcolor=dblue,
    urlcolor=dblue,
    citecolor=dblue
]{hyperref}

\newcommand{\um}{\,\mu\textrm{m}}
\newcommand{\sv}{$V^2$}
\newcommand{\mas}{\,\textrm{mas}}
\newcommand{\AU}{\,\textrm{AU}}
\newcommand{\pc}{\,\textrm{pc}}
\newcommand{\g}{\cellcolor{ratgray}}

\begin{document}

   \title{A near-infrared interferometric survey of debris-disk stars.}

   \subtitle{IV. An unbiased sample of 92 southern stars observed in H~band with \textit{VLTI}/PIONIER\thanks{Based on observations made with ESO Telescopes at the La Silla Paranal Observatory under program IDs 089.C-0365 and 090.C-0526.}}

   \author{S. Ertel \inst{1,2}
          \and
          O. Absil\inst{3}\fnmsep\thanks{F.R.S.-FNRS Research Associate}
          \and
          D. Defr\`ere \inst{4}
          \and
          J.-B. Le Bouquin \inst{1}
          \and
          J.-C. Augereau \inst{1}
          \and
          L. Marion \inst{3}
          \and
          N. Blind \inst{5}
          \and
          A. Bonsor \inst{6}
          \and
          G. Bryden \inst{7}
          \and
          J. Lebreton \inst{8,9}
          \and
          J. Milli \inst{1,2}
          }

   \institute{Univ. Grenoble Alpes, IPAG, F-38000 Grenoble, France\\
              CNRS, IPAG, F-38000 Grenoble, France
         \and
              European Southern Observatory, Alonso de Cordova 3107, Vitacura, Casilla 19001, Santiago 19, Chile\\
              \email{sertel@eso.org}
         \and
              D\'epartement d'Astrophysique, G\'eophysique et Oc\'eanographie, Universit\'e de Li\`ege, All\'ee du Six Ao\^ut 17, 4000 Li\`ege, Belgium
         \and
              Department of Astronomy, University of Arizona, 993 N. Cherry Ave, Tucson, AZ 85721, USA
         \and
              Max Planck Institute for Extraterrestrial Physics, Gie{\ss}enbachstra{\ss}e, 85741 Garching, Germany
         \and
              School of Physics, H. H. Wills Physics Laboratory, University of Bristol, Tyndall Avenue, Bristol BS8 1TL, UK
         \and 
              Jet Propulsion Laboratory, California Institute of Technology, Pasadena, CA 91109, USA
         \and
              Infrared Processing and Analysis Center, California Institute of Technology, Pasadena, CA 91125, USA 
         \and
              NASA Exoplanet Science Institute, California Institute of Technology, 770 S. Wilson Ave., Pasadena, CA 91125, USA}

   \date{April 2nd 2013}

% \abstract{}{}{}{}{} 
% 5 {} token are mandatory
 
  \abstract
   {Detecting and characterizing circumstellar dust is a way to study the architecture and evolution of planetary systems. Cold dust in debris disks only traces the outer regions. Warm and hot exozodiacal dust needs to be studied in order to trace regions close to the habitable zone.}
   {We aim to determine the prevalence and to constrain the properties of hot exozodiacal dust around nearby main-sequence stars.}
   {We searched a magnitude-limited ($H \leq 5$) sample of 92 stars for bright exozodiacal dust using our \textit{VLTI} visitor instrument PIONIER in the H~band. We derived statistics of the detection rate with respect to parameters, such as the stellar spectral type and age or the presence of a debris disk in the outer regions of the systems. We derived more robust statistics by combining our sample with the results from our \textit{CHARA}/FLUOR survey in the K~band. In addition, our spectrally dispersed data allowed us to put constraints on the emission mechanism and the dust properties in the detected systems.}
   {We find an overall detection rate of bright exozodiacal dust in the H~band of 11\% (9 out of 85 targets) and three tentative detections. The detection rate decreases from early type to late type stars and increases with the age of the host star. We do not confirm the tentative correlation between the presence of cold and hot dust found in our earlier analysis of the FLUOR sample alone. Our spectrally dispersed data suggest that either the dust is extremely hot or the emission is dominated by the scattered light in most cases. The implications of our results for the target selection of future terrestrial planet-finding missions using direct imaging are discussed.}
   {}


   \maketitle
%
%________________________________________________________________

\section{Introduction}

Debris dust around main sequence stars has often been related to the presence of colliding planetesimals left over from the planet formation process (see \citealt{kri10} and \citealt{mat14} for recent reviews). As for the most readily observable components of planetary/planetesimal systems (i.e., systems potentially composed of planets as well as small bodies orbiting the star), such debris disks are thought to give significant insight into the architecture, dynamics, and evolution of 
these systems \citep[e.g.,][]{kal05, chi09, bol12, beu14, wya03b, kuc03, kuc10, eir11, ert12a, ert12b, the12, kri13}. However, in most debris disks observed so far, the known dust is located in the outer regions of the systems, several tens of AU from the host stars and similar to -- albeit often farther than -- the location of the Kuiper belt in our solar system \citep[e.g.,][]{law09, ert11, loe12, eir13, ert14}. 

If we want to study the formation and evolution of Earth-like planets close to the habitable zone, we need to observe dust closer to this region, where it is similar to our zodiacal dust (exozodiacal dust). On 
the other 
hand, the presence of such dust around 
other stars may represent a major obstacle for future terrestrial planet-finding missions \citep{def10, def12b, rob12}. The possible presence of diffuse emission adds uncertainty to the observations. Clumpy structures in the dust distribution may point toward dynamical interaction with planets \citep{sta08}, but a clump may also be misinterpreted as an actual planet. In the recent literature, the term ``exozodiacal dust'' has been used mostly to refer to dust in the habitable zone of main sequence stars owing to the relevance of this kind of dust for detecting exo-Earths \citep[e.g.,][]{sta08, def10, mil11, rob12, ken13}.

 It is important, however, to note that the zodiacal dust in the solar system extends well beyond the borders of the habitable zone. In fact, its global radial distribution has been shown to extend continuously and with inwardly increasing surface density from a few AU from the sun down to a fraction of an AU, where it forms the F-corona \citep{kim98, hah02}. Likewise, it is 
expected that extrasolar analogs to the zodiacal dust cloud (exozodis) will span a broad range of distances, encompassing the habitable zone, but not being limited to it. In this paper, exozodiacal dust disks, or exozodis, will thus refer to any hot and/or warm dust located within a few AU from a star, down to the region where dust grains sublimate. While Kuiper belt-like debris disks emit the majority of their radiation in the far-infrared and the emission of dust in the habitable zone around a star peaks in the mid-infrared, a significant contribution from exozodiacal dust grains very close to the sublimation distance is expected to extend the emission of exozodis toward near-IR wavelengths.

Warm dust around main sequence stars showing extreme emission in the mid-infrared has been discovered photometrically or spectroscopically mostly by space-based observatories \citep[e.g.,][]{lis08, law09, lis09}, but these extreme systems are found to be rare with an occurrence rate of only $\sim1\%$. The large number of stars observed in the mid-infrared by the Wide-Field Infrared Survey Explorer (\textit{WISE}) recently allowed \citet{ken13} to detect photometrically in the mid-infrared a reasonably large sample of excesses due to exozodiacal dust for a statistical analysis. However, this is limited to the brightest excess sources with a disk-to-star flux ratio $>$$15\%$. Furthermore, these observations not only may trace warm dust, but may also be contaminated by the warm end of the emission of cold dust in the system, far away from the 
habitable zone. The first detections in the near-infrared of systems showing a small, possibly more common excess of $\sim1\%$ attributed to the presence of hot, exozodiacal dust have been reported by \citet{abs06}, \citet{dif07}, and \citet{abs08}. Detecting this dust against the bright stellar photosphere requires spatially resolved, high contrast observations, since the photometric accuracy is usually insufficient. Coronagraphy is limited by the large inner working angles of usually $>$$100\mas$ of available instruments and the small extent of the systems, because $1\AU$ at $10\pc$ corresponds to an angular distance of $100\mas$ from the star. Thus, near- or mid-infrared interferometry is the only technique available so far that allows for detecting and studying such systems.

It is, however, important to note that observations in the near-infrared are sensitive only to the hottest dust component of exozodiacal systems\footnote{One might call this component ``exo-F-corona'', but in order to avoid unnecessary complexity in the terminology we stick to ``hot exozodiacal dust'' or just ``exozodiacal dust''. We keep in mind that the border between the F-corona and the zodiacal disk is not well defined and that, depending on the focus of the respective publication, one is often just considered an extension of the other.}. On the other hand, our knowledge about our own zodiacal dust is highly biased toward its properties near the orbit of the Earth due to various observations by space-based infrared facilities \citep[e.g.,][]{sky88, ber94, pyo10}, the problems arising from observations of faint, extended emission very close to the Sun, and the naturally increased interest in this region. Furthermore, the radial distribution of the dust around other stars does not 
necessarily 
follow that of our zodiacal dust. As a consequence, connecting detections in the near-infrared to dust in the habitable zone around these stars or the zodiacal dust is not straightforward. Nonetheless, this kind of observation provides a valuable data set for investigating the dust content of the inner regions of extrasolar planetary systems.

Earlier publications mostly focused on reporting and studying single, new detections. Major advances have been made recently by two surveys on the \textit{Keck} Interferometric Nuller (KIN) in the N~band \citep{mil11} with three detections out of 22 targets and the \textit{Center for High Angular Resolution Astronomy} (\textit{CHARA}) array in the K~band \citep{abs13} with 13 detections out of 42 targets. They represent the first attempts to statistically study the incidence of exozodiacal dust depending on different parameters of the systems, such as the stellar spectral type, age, and the presence of a Kuiper belt-like debris disk. In particular, the \textit{CHARA} survey revealed first correlations, although conclusions were limited by the small sample size. These statistics tentatively suggest that the incidence of the circumstellar emission correlates with spectral type (more frequent around stars of earlier spectral type) and -- in the case of solar-type stars -- with the presence of 
significant amounts of cold dust detected through its far-infrared excess emission. 

In the present paper we extend the sample of the \textit{CHARA} survey toward the southern hemisphere and toward fainter stars using our \textit{Very Large Telescope Interferometer} (\textit{VLTI}) visitor instrument Precision Integrated Optics Near Infrared ExpeRiment (PIONIER, \citealt{lebou11}). We increase the sample size by a factor of three. In addition, the simultaneous use of four Auxiliary Telescopes (ATs) allows us to obtain closure phase measurements, directly distinguishing between uniform circumstellar emission and a companion being responsible for the excess found. Finally, our low spectral resolution data dispersed over three spectral channels in the H~band allow us to draw conclusions on the spectral slope of the excesses detected and thus on the nature of the emission and on the dust properties.

We present our sample selection in Sect.~\ref{sect_sample}. Our observing strategy and data processing are explained in Sect.~\ref{sect_data}. In Sect.~\ref{sect_results}, our results are presented and discussed. Statistics on the broad-band detection rate in the PIONIER sample are presented in Sect.~\ref{sect_stat}, and we analyze the spectrally dispersed data in more detail in Sect.~\ref{sect_spectral}. In Sect.~\ref{sect_comp_chara} we merge the \textit{VLTI}/PIONIER sample with the \textit{CHARA}/FLUOR sample and derive more robust statistics. A summary and conclusions are presented in Sect.~\ref{sect_sum_conc}.

\section{Stellar sample}
\label{sect_sample}

In this section, we present the stellar sample for our PIONIER survey and, in general, suggest guidelines for the target selection of future near-infrared interferometric search for exozodiacal dust.

\subsection{Sample creation}

In the following, we distinguish between a \emph{Kuiper belt-like debris disk} or \emph{cold dust} and \emph{exozodiacal dust} or \emph{hot dust}. The former refers to dust between several AU and a few hundred AU from the star, most likely produced by larger bodies in collisional equilibrium over the age of the system, and predominantly emitting thermally in the far-infrared. The latter refers to dust between the sublimation radius and a few AU from the star, predominantly emitting thermal radiation in the near- and mid-infrared. The main goal of our survey is to study the origin of exozodiacal dust through statistical investigation of its incidence with respect to the following properties of the observed systems:

\begin{itemize}
 \item The age of the central star. This allows us to study the evolution of the hot dust content vs.\ system age. If the dust is produced in steady state by collisions of larger bodies (at the location of the hot dust detected or somewhere else in the system and continuously drawn to this location), the dust content will decrease over time. Thus, the excess detected and with it the detection rate will also decrease \citep{wya07a, wya08, loe08}.
 \item The stellar spectral type. This can give hints to the origin and evolution of the phenomenon. The evolution of circumstellar dust is significantly affected by the stellar radiation and thus is different for stars of different luminosities. The detection rate of (cold) debris disks is well known to be decreasing from early-type toward late-type host stars\footnote{This might be due to an age bias, since the mass of debris disks decreases with age, and late type stars have a much longer main sequence lifetime than A type stars \citep{su06}.}. 
 \item The presence or absence of a debris disk in the system. The exozodiacal dust might be produced in a debris disk and dynamically drawn to the inner regions of the systems, or large bodies originating in the outer belt might be transported to the inner system and produce the dust through collisions there \citep{bon12, bon13b}. In this case, a clear correlation between the presence of the cold and hot dust would be expected.
\end{itemize}

To achieve these goals, a carefully selected target list is mandatory in order to avoid selection biases. We consider a list of debris disk detections and non-detections in the far-infrared available to us (by April 2012). These data come from three sources:
\begin{itemize}
 \item A list of all stars observed by the \textit{Spitzer} Space Telescope in the context of debris disk programs. Published fluxes at $24\um$ and $70\um$ where available (see excess references in Table~\ref{tab_targets}), as well as archive data, were considered. The archive data were checked for $70\um$ excess by predicting the flux at $70\um$ from that at $24\um$ using the Rayleigh-Jeans law. This method was checked considering published detections, and the results were found to be in good agreement. If only detections were published for a survey and if a star observed was not included in this publication, it is assumed that indeed no excess was detected.
 \item The results from the \textit{Herschel}/DUNES survey \citep{eir10, eir13} including all detections and non-detections of excesses.
 \item A list of preliminary non-detections of excess (G. Kennedy, personal communication) from the \textit{Herschel}/DEBRIS survey \citep{mat10}.
 \item A reduction of the data for an incomplete list of targets observed by other \textit{Herschel} debris disk programs taken from the archive.
\end{itemize}

According to this information, we distinguish between \emph{debris stars} (stars with a debris disk detected) and \emph{control stars} (stars that have been searched for a debris disk, but none was detected). In case of controversial information, \textit{Herschel} data are considered more reliable than \textit{Spitzer} data due to the higher angular resolution and usually higher sensitivity and DUNES data are considered more reliable than DEBRIS data due to typically higher sensitivity and because the DUNES survey results have been published already.

Further refinement is needed to remove targets unsuitable for interferometric observations. This is the case if a target is too faint for our high-accuracy observations (i.e., minimizing the piston noise by scanning the fringes as fast as possible, but with enough flux not to reach the photon-noise regime). Sources as faint as $H = 4$ can comfortably be observed in this mode at typical conditions using PIONIER with the ATs. The regime of $4 < H < 5$ is accessible under good atmospheric conditions. Targets fainter than $H = 5$ have been removed. Furthermore, very bright stars cannot be observed because they saturate the detector. This is the case, for example, for the otherwise very obvious target Fomalhaut \citep{leb13}. Binary companions within the interferometric field of view ($\sim$$400\,\textrm{mas}$ full width at half maximum in H~band) prevent us from detecting weak, extended circumstellar emission and even light from close companions outside the field of view may enter the optical path in case of bad 
seeing. Thus, all known binary systems with angular separation $< 5''$ are removed from the samples, using the catalogs of \citet{pou04}, \citet{egg08}, \citet{rag10}, and \citet{dom02}.

In addition, stars with unusually large linear diameters for main sequence stars of a given spectral type are removed. This is a signpost of post main sequence evolution which might result in physical phenomena, such as outflows that would be misinterpreted as exozodiacal dust. The stellar angular diameter $\theta_{V-K}$ is estimated from V and K colors using the surface brightness relation \citet{ker04}:
\begin{equation}
 \label{eq_diam}
 \log\theta_{V-K} = 0.2753 \left(V-K\right) + 0.5175 - 0.2V~.
\end{equation}
The linear diameter $D_\star = \theta_{V-K} / d$ is computed from that value using the Hipparcos \citep{per97} distances $d$ of the stars. V~band magnitudes are taken from \cite{kha09}. We consider K~band magnitudes from the Two Micron All Sky Survey (2MASS, \citealt{skr06}) where the data quality is sufficient. For targets with 2MASS quality flag $\neq A$, for example, due to saturation, data collected by \citet{gez93} (updated in 1999) are used, averaging the listed measurements and considering typical uncertainties quoted in the references. If no data are available from \citet{gez93}, the 2MASS data and appropriate uncertainties are used. To identify stars with large diameters, our estimated linear diameters $D_\star$ are compared to theoretical values $D_{\star,\textrm{th}}$ from Allen's Astrophysical Quantities \citep{cox00}. Stars above an empirically defined threshold $D_{\star\text{cut}}$ are removed:
\begin{equation}
 \label{eq_thres}
 D_{\star,\text{cut}} = D_{\star,\text{th}} + a \sigma_\textrm{D} \frac{D_{\star,\text{th}}}{\left\langle D_{\star,\text{th}}\right\rangle}~.
\end{equation}
Here, $D_{\star,\text{th}}$ is the theoretical linear diameter derived from a quadratic fit to theoretical values vs.\ spectral types, $\sigma_\textrm{D}$ is derived from the scatter of the diameters for a given spectral type, $\left\langle D_{\star,\text{th}}\right\rangle$ is the arithmetic mean of $D_{\star,\text{th}}$ of all stars, and $a$ is chosen to be 1.5. Available ages of the stars removed have been checked. Nearly all removed stars have ages comparable to or larger than their main sequence life time, indicating that the method is successful.

To obtain comparable results for debris stars and control stars, the two samples have to be as similar as possible with respect to their distribution in spectral type, brightness, and observing conditions (i.e., sensitivity). This can be achieved by selecting and observing pairs of debris stars and most similar control stars directly after each other (which ideally means red very similar conditions), where possible.

This results in an all sky sample of targets. For each PIONIER run the suitable targets are selected from this sample with special attention on observing a sample of stars that is balanced between the three spectral type bins of A type stars, F type stars, and G and K type stars. Only very few M type stars remain in the sample due to the brightness limitations. They are thus not considered for any spectral type bin.

\subsection{Properties of the observed targets}
\label{sect_target_prop}

A list of stellar parameters and near-infrared photometry of our observed targets is given in Table~\ref{tab_targets}. Angular diameters $\theta_{V-K}$ are computed following Sect.~\ref{sect_sample}. Age estimates were collected from the \texttt{VizieR} data base\footnote{http://vizier.u-strasbg.fr/viz-bin/VizieR}. The mean logarithmic ages are computed from all independent estimates available. Exceptions have been made for $\beta\,\textrm{Pic}$ and HD\,172555, which are well-established members of the $\beta\,\textrm{Pic}$ moving group \citep{zuc01}. Here, we consider the latest estimates for the age of this group \citep{bin14}. For two targets, HD\,141891 and HD\,128898, no age estimates were found. HD\,141891 is an old F-type star for which we will see later that even a non-detection is relevant for the statistics of excess detection vs.\ age (Sect.~\ref{sect_stat_age}). We estimate the age from the bolometric and X-ray luminosity \citep{and12, schm04} following \citet{mam08}. HD\,128898 is an A type star 
without hot excess as we show in Sect.~\ref{sect_results}. For this age bin, the inclusion or not of one more non-detection does not significantly affect our statistics. Thus, we exclude this target from the age statistics. The age values are listed in Table~\ref{tab_targets}.

\section{Data acquisition and processing}
\label{sect_data}

\subsection{Detection strategy}
\label{sect_detectionstrategy}

\begin{figure}
 \centering
 \includegraphics[angle=0,width=\linewidth]{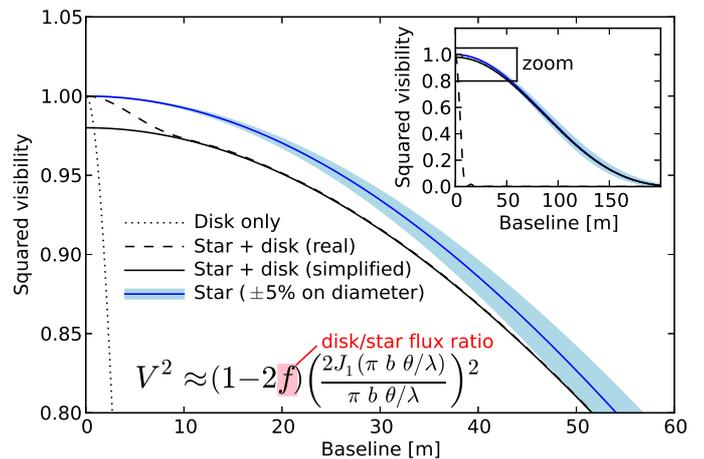}
 \caption{Illustration of our detection strategy following \citet{dif07}. For the `real', dashed curve we assume a uniform disk for both the star and the flux distribution from the exozodiacal dust and a disk-to-star flux ratio of $f=0.01$, while for the `simplified', solid curve we use the same assumptions but the approximation following the equation in the figure. Diameters of the star and (face-on) disk have been chosen to 2.5\,mas (about an A-type star at 10\,pc) and 500\,mas (5\,AU at 10\,pc), but exact numbers are not relevant for the illustration of our detection strategy. For details see Sect.~\ref{sect_detectionstrategy}.}
 \label{fig_detectionstrategy}
\end{figure}

When it comes to the detection of faint, circumstellar excess emission, the strength of (near-) infrared interferometry is the ability to spatially resolve this emission and thus to spatially disentangle it from the much brighter stellar emission. Therefore we follow the approach first presented by \citet{dif07} and briefly summarized here. When observing at small baselines of up to a few tens of meters, a nearby star is nearly unresolved. This minimizes the effect of its uncertain diameter on the prediction of its squared visibility (\sv). At the same time, an extended circumstellar emission is ideally fully resolved. This will result in a drop in \sv\ compared to the purely stellar \sv, because it adds incoherent flux. This represents the core of our detection strategy and is illustrated in Fig.~\ref{fig_detectionstrategy}. Measurements on a limited range of baselines, however, do not allow one to directly distinguish between a faint companion and a circumstellar disk. The availability of closure phase 
data 
allows distinguishing between azimuthally symmetric emission from a circumstellar disk and highly asymmetric emission from a companion \citep{lebou12a, mar14}.

\subsection{Overview}

In this section, we describe the acquisition and processing of the data from the observations to the measurement of the disk-to star flux ratio in case of detected circumstellar excess emission. This is a complex, multi-step process with some decisions in earlier steps being motivated by the requirements during later steps. Thus, we give a quick overview here first before discussing each step in detail in the following sections:

\begin{itemize}
 \item \textbf{Observation:} We measure the squared visibility of our targets on six baselines (4 telescopes) simultaneously. Observations of one target are interrupted by identical observations of calibrators. For details, see Sect.~\ref{sect_obs}.
 \item \textbf{Data reduction} is carried out using the dedicated script of the PIONIER data reduction pipeline. For details, see Sect.~\ref{sect_red_cal}.
 \item \textbf{Calibration} of the measured squared visibilities is done with the dedicated script of the PIONIER data reduction pipeline. From the observed sequences of calibrators (CAL) and science targets (SCI) we select CAL-SCI or SCI-CAL pairs observed directly after each other to compare their squared visibilities. Several effects such as chromaticism have to be characterized and considered in detail to achieve the accuracy we aim for with our survey. For details, see Sect.~\ref{sect_red_cal}.
 \item \textbf{Analysis of closure phase data}  to reject targets with companions. See Sect.~\ref{sect_closure_phase} for details.
 \item \textbf{Measuring the excess} with the high accuracy needed to detect possible excesses requires the combination of all measurements of one target in order to achieve a high cumulative accuracy. Therefore, we use a simple model of the instrumental response to extended emission. See Sect.~\ref{sect_model_fitting} for details.
\end{itemize}

\subsection{Observation}
\label{sect_obs}
Observations were carried out in H~band in two runs each in P89 (Apr.\ 2012 and Jul.\ 2012) and P90 (Oct.\ 2012 and Dec.\ 2012), each run consisting of three consecutive observing nights. In total, 92 stars were observed. An observing log of all nights can be found in Table~\ref{tab_obslog}.

\addtocounter{table}{2}
\begin{table*}
\caption{Summary of \textit{VLTI}/PIONIER observations}             
\label{tab_obslog}      
\centering          
\begin{tabular*}{1.0\textwidth}{l@{\extracolsep{\fill}}ccccl} 
\toprule
  Run           & Night      & \# targets & Seeing [$''$]   & $t_0$ [ms]     & Condition notes \\
\midrule
  089.C-0365(A) & 2012-04-27 &  0         & --              & --             & Night lost due to bad weather  \\
  089.C-0365(A) & 2012-04-28 & 10         & 1.2 (0.7--1.8)  & 2.0 (1.5--2.1) & Average conditions \\
  089.C-0365(A) & 2012-04-29 & 11         & 0.8 (0.6--1.8)  & 4.5 (1.5--6.2) & Good conditions \\
  089.C-0365(B) & 2012-07-23 &  9         & 1.0 (0.7--1.7)  & 3.0 (2.0--4.0) & Good conditions \\
  089.C-0365(B) & 2012-07-24 &  9         & 0.7 (0.5--1.0)  & 3.5 (2.5--5.0) & Good conditions \\
  089.C-0365(B) & 2012-07-25 & 13         & 0.9 (0.6--1.7)  & 2.5 (1.5--4.0) & Good conditions \\
  090.C-0526(A) & 2012-10-14 & 11         & 0.8 (0.6--1.3)  & 2.0 (1.5--3.0) & Good conditions \\
  090.C-0526(A) & 2012-10-15 &  6         & 1.3 (0.8--2.4)  & 1.5 (1.0--2.5) & Bad conditions, tech. loss \\
  090.C-0526(A) & 2012-10-16 & 11         & 0.9 (0.6--1.9)  & 1.5 (1.0--3.0) & Good conditions \\
  090.C-0526(B) & 2012-12-15 &  6         & 0.9 (0.5--1.5)  & 5.0 (3.0--7.0) & Good conditions, tech. loss \\
  090.C-0526(B) & 2012-12-16 &  8         & 1.2 (0.7--2.1)  & 3.0 (2.0--5.0) & Bad conditions \\
  090.C-0526(B) & 2012-12-17 &  8         & 0.8 (0.6--2.0)  & 4.0 (2.0--6.0) & Average conditions \\
\bottomrule                
\end{tabular*}
\tablefoot{The seeing is measured as DIMM seeing in the visible. The quantity $t_0$ the nominal coherence time. The first value in both columns gives the typical value for the night, and the values in parentheses the range observed. In addition, the condition notes column gives an evaluation of the conditions based on instrument performance and data quality, as the nominal values of $r_0$ and $t_0$ are not always fully correlated with the actual performance of the observations (e.g., due to local turbulences at the ATs not measured by the DIMM seeing). The sum of the number of targets is larger than the total number of targets observed (92), as some targets have been re-observed due to limited data quality or the observations have been carried out over two nights owing to timing constraints.}
\end{table*}

We used the four 1.8-m ATs to obtain six visibility measurements simultaneously. The most compact array configuration available at the \textit{VLTI} with baselines between 11\,m and 36\,m was selected. The detector read-out mode was set to FOWLER with the \textit{SMALL} dispersion (three spectral channels) and only outputs A and C read in order to speed up the readout. The number of steps read in one scan (\textit{NDREAD}) was 1024. See \citealt{lebou11} for a description of the available modes and their effects. This setup was used for all observations (besides a few with slightly different setups tried to optimize the strategy). Instead of adjusting the instrument setup accounting for faint targets or bad conditions, we selected brighter targets in case of worse conditions and vice versa. This guaranteed a homogeneous observing setup for the whole sample.

Three calibrators were selected from \citet{mer05} for each science target, typically within $10\degr$ to minimize the effects of the pupil rotation (i.e., position on sky, \citealt{lebou12}). Additional criteria were similar H~band brightness and small angular diameter. Most of the targets have been observed in a sequence of CAL1--SCI--CAL2--SCI--CAL3--SCI--CAL1 and where possible, the corresponding debris or control star has been observed directly afterward.

\subsection{Data reduction and calibration}
\label{sect_red_cal}

In this section we describe the conversion of raw observations into calibrated interferometric observables (\sv\ and closure phase). Data reduction used the standard PIONIER pipeline \texttt{pndrs} version 2.51 \citep{lebou11}. The five consecutive files composing an observing block, either SCI or CAL, were averaged together to increase S/N and reduce the amount of data to be dealt with.

\subsubsection{Nightly-based, global calibration}

As explained by \citet{lebou12}, we identified that the major source of instrumental instability in the data is linked to the pupil rotation inside the \textit{VLTI} optical train. At first order, the instrumental contrast (transfer function, TF, i.e., the measured but not calibrated \sv\ of a point source given instrumental and atmospheric effects) is described by
\begin{equation}
  C = a + b \cos(alt + az - 18\degr) \label{eq_TF_polar}
\end{equation}
where $alt$ and $az$ are the actual elevation and azimuth.

For typical values of $a\approx0.7$ and $b\approx0.1$, this means that the difference of $alt+az$ between the observations of SCI and CAL should be smaller than $2\degr$ for this effect to be less than the desired level of accuracy. Since the density of calibrators in the sky is not sufficient, we correct for this effect by implementing a global analysis of each night, before the classical SCI/CAL calibration.

We first fit Eq.~\ref{eq_TF_polar} to all calibrators of the night in order to determine the parameters $a$ and $b$. This fit is well constrained because we typically gather about 35 observations of calibrators during a single night, spread all over the sky. Then we use $a$ and $b$ to correct all the observations of the night, which is all SCI but also all CAL. After this correction, the average level of the instrumental response within a CAL-SCI-...-CAL sequence is generally not unity, because Eq.~\ref{eq_TF_polar} suffers from idealization. Nevertheless, this strategy successfully removes any spurious trend that could be associated to the global pointing dependency. Consequently, the typical scatter among the four observations of calibrators within the sequence is reduced to two to three percent.

\subsubsection{SCI/CAL calibration}

The goal of this second step is to correct for the instrumental response within the CAL-SCI-...-CAL sequence. To ease the implementation of the subsequent bootstrapping analysis (see Sec.3.7), we do not average all the CAL observations into a single value of the instrumental response. Instead, we calibrate each SCI individually by pairing it either with the preceding or the following CAL. Calibrators with low S/N or with a clear closure phase signal are rejected. This accounts for about 1\% of all calibrators observed. Furthermore, the same CAL observation is never used to calibrate two SCI, in order to minimize the correlation. Finally, where this is possible, only one of the two observations on the same calibrator in one sequence is used to maximize the number of different calibrators (ideally 3). For each object we gather a total of three calibrated observations, each of them calibrated by a different calibrator. Each observation contains six squared visibilities 
measurements (\sv) and four 
closure-phase measurements, each of them dispersed over three spectral channels with central wavelengths of $1.59\um$, $1.68\um$, and $1.77\um$.

\subsection{Assessment of systematic chromatic effects}
\label{sect_chromaticism}

A conservative estimate of the chromaticism of PIONIER was given by \citet{def12}. Given the precision intended for our survey, we have to quantify this effect. This is particularly important since most calibrators from \citet{mer05} are K giants, while our science targets are distributed over spectral types A to K. We describe in Appendix~\ref{app_chromat} the detailed analysis of the chromaticism carried out based on our data set. We empirically derive a correction for the chromaticism, but find that the effect of this correction on our results is negligible. In a few cases the correction derived is much larger but obviously erroneous (mostly due to noisy data). Thus, we do not apply the correction to the data set used for the further analysis presented in this paper, as the potential gain is minor, while there is the risk of a failure of the correction for some targets, which would result in a significant degradation of the data quality. Instead, we consider a conservative, systematic calibration 
uncertainty 
of $1\times10^{-3}$ on the squared visibility measurements. This uncertainty is correlated among all data and results in an uncertainty on the star-to-disk flux ratio of $5\times10^{-4}$.

\begin{figure*}
 \centering
 \includegraphics[angle=0,width=\linewidth]{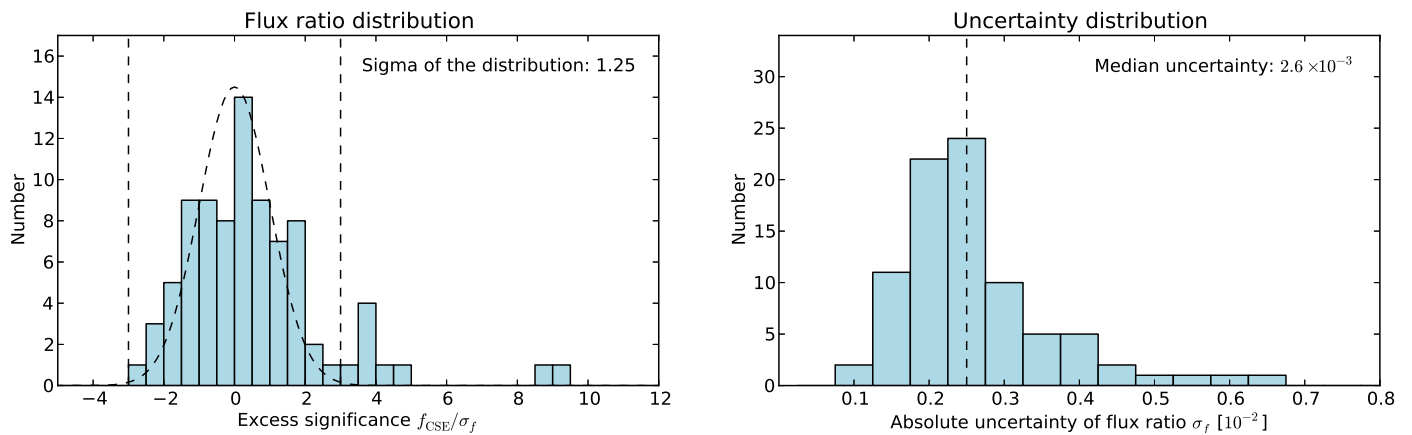}
 \caption{Excess distribution (\emph{left}) and distribution of uncertainties on the disk-to-star flux ratio (\emph{right}). The Gaussian overplotted on the excess distribution has a width of $\sigma = 1$ and is used to guide the eye and illustrates that the data are consistent with this ideal behavior. Vertical dashed lines are plotted at $f_\textrm{CSE}/\sigma_f = -3$ and $f_\textrm{CSE}/\sigma_f = +3$ for the excess distribution and at the median uncertainty ($2.5\times10^{-3}$) for the uncertainty distribution.}
 \label{fig_hist_excess}
\end{figure*}

\subsection{Analysis of closure phase data}
\label{sect_closure_phase}

The closure-phase data obtained in the context of the present project constitute a valuable sample for the search for unknown, faint companions around nearby main sequence stars. We analyze the closure phase data in detail in another paper \citep{mar14}
for our further analysis. Here, we rely on the results of this work and only discard the systems in which companions have been detected.

Five of the 92 targets observed -- HD\,4150, HD\,16555, HD\,29388, HD\,202730, and HD\,224392 -- show a closure phase signal that can be attributed to the presence of a previously unknown stellar companion and thus have to be removed from the subsequent analysis. In addition, a companion around HD\,15798 ($\sigma$\,Cet) has been detected by \citet{tok14} using speckle interferometry. Finally, we reject HD\,23249 ($\delta$\,Eri) because of potential post-main sequence evolution (Sect.~\ref{sect_specific}), which leaves us with 85 stars.

\subsection{Fitting of exozodiacal dust models}
\label{sect_model_fitting}

In the present section, we describe the fitting strategy used to combine all \sv\ data of a given object in order to derive a disk-to-star flux ratio (hereafter flux ratio). The flux ratio does not depend significantly on the assumed disk geometry as shown in previous studies (e.g., \citealt{abs09, def11}). We consider a model consisting of a limb-darkened photosphere surrounded by a uniform emission filling the entire field of view of PIONIER on the ATs (see analytical expression in Fig.~\ref{fig_detectionstrategy}). Under typical seeing conditions, this field of view can be approximated by a Gaussian profile with a full width at half maximum of 400 mas \citep{abs11}. The visibility expected from a limb-darkened photosphere is estimated according to \citet{han74} using the linear H~band limb-darkening coefficients of \citet{cla95}. We estimate the visibility for the whole bandwidth of each spectral channel, considering the actual spectrum of the star using tabulated H~band spectra from \citet{pic98} and 
the spectral transmission of the PIONIER instrument. The estimated squared visibilities are then compared to the measurements and the flux ratio for each data set derived.

The computation is performed by a set of IDL routines developed initially for \textit{CHARA} observations \citep{abs06} and adapted later for more telescopes \citep{def11}. To derive the value and uncertainty of the flux ratio for each target, we use a bootstrapping algorithm with each individual fit to the data performed using a Levenberg-Marquardt least-squares minimization \citep{mar09}. This means that the individual uncertainties on the data points are not considered directly in the estimate of the uncertainty of the flux ratio, but rather their scatter. In addition, a systematic uncertainty on each data point as caused by the uncertain diameter of the calibrator is considered. Finally, a systematic uncertainty of $5\times10^{-4}$ due to the chromaticism is added to the flux ratio derived (Sect.~\ref{sect_chromaticism}).

For the bootstrapping, we investigate several possible correlations among the data. These could be present, for example, among the different spectral channels in which data have been obtained simultaneously, the baselines sharing one telescope, or data that were obtained on all six baselines simultaneously in one OB, since these share the same calibrator. We fit the whole sample several times, each time assuming one of these correlations to be the dominant one. The level of correlation left in the data after the fit is estimated by the width of the distribution of excesses weighted by their uncertainties. Such a histogram should ideally have a Gaussian shape with a standard deviation of one if there is no detection at all among the sample. Fewer than 1\% of the targets should have a flux ratio $<-3\sigma$, while some significant detections should show up 
with a flux ratio $>3\sigma$. A smaller scatter suggests an overestimation of the correlation, while larger scatter and a significant number of targets 
with flux ratio $<-3\
sigma$ suggest an underestimation.

We find that the correlation among the three spectral channels dominates and that all other correlations can be neglected. This is expected because the spectral channels share the same \textit{VLTI} beams and so have the same polarization behavior, which is the dominant source of systematic error. Moreover, they share the same piston statistics, which is the dominant source of statistical noise. Figure~\ref{fig_hist_excess} shows the histogram of the significance of the flux ratios for our sample in this case, as well as the sensitivities reached ($1\sigma$).

\section{Results}
\label{sect_results}

\begin{table*}
\caption{Detections (marked in gray) and non-detections of extended emission and closure-phase signal}
\label{tab_detections}      
\centering          
\begin{tabular}{ccccccccccccccc} 
\toprule
HD      & $f_\textrm{CSE}$ [\%] & $\sigma_f$ [\%] & $\chi_f$ & $\chi^2_\textrm{red}$ & $V^2$    &  Comp.  & ~~ & HD     & $f_\textrm{CSE}$ [\%] & $\sigma_f$ [\%] & $\chi_f$ & $\chi^2_\textrm{red}$ & $V^2$    &  Comp. \\
\midrule
142     & -0.69  & 0.26   & -2.61  & 1.05   & no     & no    & & 91324    & 0.14   & 0.17   & 0.82   & 0.38   & no     & no    \\
1581    & -0.21  & 0.31   & -0.67  & 1.00   & no     & no    & & 99211    & 0.35   & 0.22   & 1.60   & 0.75   & no     & no    \\
\g2262  & \g0.67 & \g0.18 & \g3.72 & \g0.51 & \g yes & \g no & & 102365   & 0.43   & 0.23   & 1.91   & 0.66   & no     & no    \\
3302    & 0.36   & 0.26   & 1.36   & 1.22   & no     & no    & & \g104731 & \g0.52 & \g0.14 & \g3.73 & \g0.55 & \g yes & \g no \\
3823    & 0.25   & 0.22   & 1.16   & 0.78   & no     & no    & & \g108767 & \g0.57 & \g0.15 & \g3.73 & \g0.54 & \g yes & \g no \\
4150    & 2.57   & 0.49   & 5.22   & 2.83   & yes    & yes   & & 109787   & -0.32  & 0.20   & -1.61  & 0.73   & no     & no    \\
7570    & -0.36  & 0.26   & -1.36  & 1.17   & no     & no    & & 115617   & 0.09   & 0.23   & 0.40   & 0.40   & no     & no    \\
\g7788  & \g1.43 & \g0.17 & \g8.53 & \g0.47 & \g yes & \g no & & 120136   & -0.21  & 0.22   & -0.96  & 0.65   & no     & no    \\
10647   & -0.08  & 0.26   & -0.31  & 0.81   & no     & no    & & 128898   & 0.15   & 0.22   & 0.69   & 0.75   & no     & no    \\
11171   & -0.06  & 0.42   & -0.14  & 1.30   & no     & no    & & 129502   & -0.04  & 0.14   & -0.29  & 0.38   & no     & no    \\
\g14412 & \g0.96 & \g0.21 & \g4.66 & \g0.74 & \g yes & \g no & & 130109   & -0.41  & 0.43   & -0.96  & 1.75   & no     & no    \\
15008   & 0.56   & 0.32   & 1.73   & 1.22   & no     & no    & & 134083   & -0.56  & 0.47   & -1.18  & 2.64   & no     & no    \\
15798   & 3.03   & 0.34   & 8.82   & 0.68   & yes   & no$^1$ & & 135379   & 0.18   & 0.37   & 0.48   & 1.01   & no     & no    \\
16555   & 40.55  & 2.45   & 16.55  & \dots  & yes    & yes   & & 136202   & -1.56  & 0.64   & -2.46  & 3.03   & no     & no    \\
17051   & -0.24  & 0.23   & -1.06  & 0.96   & no     & no    & & 139664   & 0.11   & 0.19   & 0.59   & 0.62   & no     & no    \\
17925   & -0.05  & 0.23   & -0.22  & 0.70   & no     & no    & & 141891   & -0.1   & 0.20   & -0.50  & 0.64   & no     & no    \\
19107   & 0.47   & 0.21   & 2.28   & 0.49   & no     & no    & & 149661   & 0.14   & 0.22   & 0.65   & 1.70   & no     & no    \\
20766   & 0.08   & 0.26   & 0.30   & 1.18   & no     & no    & & 152391   & 0.06   & 0.18   & 0.34   & 0.58   & no     & no    \\
\g20794 & \g1.64 & \g0.37 & \g4.46 & \g1.58 & \g yes & \g no & & 160032   & 0.01   & 0.11   & 0.09   & 0.28   & no     & no    \\
20807   & -0.05  & 0.53   & -0.09  & 3.74   & no     & no    & & 160915   & 0.19   & 0.28   & 0.67   & 0.67   & no     & no    \\
22001   & 0.3    & 0.20   & 1.53   & 0.71   & no     & no    & & 164259   & -0.18  & 0.19   & -0.96  & 0.62   & no     & no    \\
23249   & 2.44   & 0.37   & 6.65   & 2.17   & yes    & no    & & 165777   & 0.46   & 0.28   & 1.62   & 0.98   & no     & no    \\
25457   & -0.07  & 0.14   & -0.50  & 0.37   & no     & no    & & 172555   & 0.55   & 0.25   & 2.16   & 1.09   & no     & no    \\
\g28355 & \g0.88 & \g0.09 & \g9.33 & \g0.36 & \g yes & \g no & & 178253   & 0.15   & 0.36   & 0.41   & 1.68   & no     & no    \\
29388   & 3.84   & 0.46   & 8.35   & 3.02   & yes    & yes   & & 182572   & 0.09   & 0.13   & 0.69   & 0.37   & no     & no    \\
30495   & -0.14  & 0.21   & -0.68  & 0.46   & no     & no    & & 188228   & 0.53   & 0.27   & 1.99   & 0.82   & no     & no    \\
31295   & 0.21   & 0.15   & 1.41   & 0.54   & no     & no    & & 192425   & -0.31  & 0.25   & -1.26  & 1.03   & no     & no    \\
31925   & 0.41   & 0.22   & 1.90   & 0.63   & no     & no    & & 195627   & 0.05   & 0.52   & 0.10   & 2.87   & no     & no    \\
33111   & 0      & 0.41   & 0.00   & 2.00   & no     & no    & & 197157   & 0.35   & 0.30   & 1.15   & 1.03   & no     & no    \\
33262   & 0.27   & 0.21   & 1.31   & 0.58   & no     & no    & & 197692   & -0.14  & 0.20   & -0.71  & 0.48   & no     & no    \\
34721   & -0.36  & 0.21   & -1.75  & 0.61   & no     & no    & & 202730   & 29.56  & 9.96   & 2.97   & \dots  & no     & yes   \\
38858   & -0.69  & 0.29   & -2.34  & 1.10   & no     & no    & & 203608   & -0.74  & 0.34   & -2.15  & 1.52   & no     & no    \\
\g39060 & \g0.88 & \g0.23 & \g3.90 & \g0.49 & \g yes & \g no & & 206860   & 0.21   & 0.30   & 0.69   & 1.24   & no     & no    \\
40307   & -0.34  & 0.24   & -1.44  & 1.06   & no     & no    & & 207129   & 0.13   & 0.13   & 1.00   & 0.26   & no     & no    \\
43162   & 0.4    & 0.21   & 1.94   & 0.63   & no     & no    & & 210049   & 0.18   & 0.38   & 0.47   & 2.67   & no     & no    \\
45184   & 0.42   & 0.15   & 2.83   & 0.53   & no     & no    & & 210277   & -0.41  & 0.31   & -1.31  & 1.85   & no     & no    \\
53705   & 0.08   & 0.23   & 0.35   & 0.67   & no     & no    & & \g210302 & \g0.83 & \g0.25 & \g3.39 & \g1.07 & \g yes & \g no \\
56537   & -0.42  & 0.25   & -1.68  & 0.98   & no     & no    & & 210418   & -0.43  & 0.29   & -1.46  & 1.17   & no     & no    \\
69830   & 0.04   & 0.26   & 0.15   & 1.01   & no     & no    & & 213845   & -0.43  & 0.24   & -1.81  & 0.82   & no     & no    \\
71155   & 0.09   & 0.25   & 0.35   & 1.16   & no     & no    & & 214953   & -0.15  & 0.22   & -0.69  & 0.71   & no     & no    \\
72673   & 0      & 0.33   & 0.00   & 1.89   & no     & no    & & 215648   & -0.21  & 0.22   & -0.96  & 0.76   & no     & no    \\
76151   & 0.16   & 0.28   & 0.56   & 1.38   & no     & no    & & 215789   & -0.2   & 0.26   & -0.78  & 0.68   & no     & no    \\
76932   & -0.04  & 0.42   & -0.09  & 2.79   & no     & no    & & 216435   & -0.35  & 0.27   & -1.27  & 1.14   & no     & no    \\
82434   & 0.39   & 0.58   & 0.68   & 3.72   & no     & no    & & 219482   & 0.2    & 0.17   & 1.19   & 0.53   & no     & no    \\
88955   & -0.25  & 0.25   & -1.02  & 1.00   & no     & no    & & 219571   & 0.09   & 0.27   & 0.33   & 0.78   & no     & no    \\
90132   & -0.74  & 0.41   & -1.79  & 1.80   & no     & no    & & 224392   & 1.74   & 0.27   & 6.34   & 1.15   & yes    & yes   \\
\bottomrule
\end{tabular}
\tablefoot{Detections, i.e., stars with significant excess but without a companion following \citet{mar14}, are marked in gray. The quantity $f_\textrm{CSE}$ is the disk-to-star flux ratio. The quantity $\sigma_f$ is the $1\sigma$ uncertainty on that measurement, $\chi_\textrm{f} = f_\textrm{CSE} / \sigma_f$ gives the significance of the detection, and $\chi^2_\textrm{red}$ gives the reduced $\chi^2$ of the fit of our model to the data. The columns named `$V^2$' and `Comp.' include notes on whether there is a significant detection of extended emission or of closure-phase signal, respectively.\\
$^1$~See notes in Sect.~\ref{sect_specific}.}\\
\end{table*}

We observed a sample of 92 stars looking for faint, extended near-infrared excess. Five targets show significant closure phase signal, which is indicative of a companion \citep{mar14}, which makes the targets useless for our analysis. One target -- HD\,15798 ($\sigma$\,Cet) -- needed to be rejected because it has a companion not detected by our analysis, and another target -- HD\,23249, $\delta$\,Eri -- needed to be rejected because it is probably a post-main-sequence star (Sect.~\ref{sect_specific}). This leaves us with a sample of 85 stars that can be used for the subsequent analysis.

For this sample, the median $1\sigma$ accuracy of the measurement of the disk-to-star flux ratio is $2.6\times10^{-3}$, i.e. 0.26\%. As shown in Fig.~\ref{fig_hist_excess}, the excess distribution is consistent with a Gaussian with $\sigma = 1$. Thus, we consider an excess to be significant if the flux ratio is higher than its $3\sigma$ uncertainty. Using this criterion, we find that 9 out of 85 stars ($10.6^{+4.3}_{-2.5}\%$) show a significant visibility drop in broad band (Table~\ref{tab_detections}). We interpret this as faint, extended circumstellar emission, which we attribute to the presence of exozodiacal dust.

In the following, we first briefly discuss a few peculiar targets (Sect.~\ref{sect_specific}). Afterward, we statistically analyze the broad band detection rate (Sect.~\ref{sect_stat}) and discuss the spectrally dispersed data obtained (Sect.~\ref{sect_spectral}).

\subsection{Notes on specific targets}
\label{sect_specific}

\textbf{HD\,15798 ($\sigma$\,Cet)} Has a significant signal in the $V^2$ but no significant closure-phase signal. This would identify the star as an excess star in our sample. However, \citet{tok14} detected a companion to this star at a separation of 210\,mas using speckle interferometry. This separation is too large for the fringe patterns of the two stars to overlap, so that the companion could not be detected by our observations. However, it is expected to contribute some incoherent flux to the observations that may well be responsible for the detected $V^2$ drop. As a consequence, the star has to be rejected from our subsequent analysis because no conclusion on the potential presence of an excess is possible. This example illustrates that even the availability of closure phase data does not completely rule out the possibility of false excess detections due to any unknown companions to our targets.

\textbf{HD\,23249 ($\delta$\,Eri)} has been observed with \textit{VLTI}/VINCI in K~band between October~2001 and February~2003 \citep{the05}. These data show no evidence of circumstellar excess, in contrast to our PIONIER data in H~band taken in October 2012 (Table~\ref{tab_detections}). The sensitivity of the two observations is comparable, ruling this out as a source of the discrepancy. That the excess has been detected in H~band but not in K~band suggests that a specific temperature of the dust cannot be responsible for the discrepancy either, since a black body of any temperature lower than the sublimation temperature would peak longward of the H~band, implying a rising flux ratio toward the K~band. Thus, assuming that the emission originates in circumstellar dust, the only explanation known would be an increase in the excess between the two 
observations. \citet{the05} also determined the evolutionary state of HD\,23249 to be at the end of the subgiant phase. Thus, our 
observations could trace emission 
stemming from physical processes related to that evolutionary state rather than to circumstellar dust. As a consequence, we reject this target from subsequent analysis. We note that this target is similar to $\kappa$\,CrB, which has been observed as part of the FLUOR sample and has been rejected for the same reason (\citet{abs13}.

\textbf{HD\,39060 ($\beta$\,Pic)} had been observed extensively with PIONIER before, resulting in an excess detection \citep{def12}. We observed the target again as part of our unbiased sample and confirm the detection. At the same time, we did not detect any significant variation of the excess between the observations (December 2010 to November 2011 vs.\ October 2012). However, the excess in our survey data is only detected at $3.9\sigma$. At this level of accuracy, we can only rule out variability over $\sim$80\% of the total flux. We confirm the flat spectral slope of the flux ratio discussed in detail by \citet{def12}.

\textbf{HD\,69830 and HD\,172555} have strong excess emission detected in the mid infrared (HD\,69830: \citealt{bei05}, HD\,172555: \citealt{schu05, che06}). We do not detect significant broad band excess in H~band around these two stars. HD\,69830 was also observed with FLUOR and no K~band excess was detected either \citep{abs13}. Given the large number of detections in our samples and the strong mid-infrared excess found for these targets, the non-detections are surprising. For HD\,172555 we find a tentative excess based on significant excess in only the long wavelength channel and on a spectral slope of the flux ratio that suggest increasing excess toward longer wavelengths. We discuss this in more detail in Sect.~\ref{sect_spectral}. Because this can only be considered a tentative detection, we do not include this potential excess in our statistical analysis. No significant far-infrared excess has been detected around HD\,69830 \citep{eir13}. Thus, this star is not counted as a 
cold dust host star in our statistical 
analysis. 
The dust in this system is located at $\sim1\,\textrm{AU}$ from the star \citep{smi09}. It is doubtful whether this dust can be produced at this location in an equilibrium collisional cascade of larger bodies over the age of the system \citep{lis07}. No significant amounts of cold dust are found. This would qualify the dust in this system as exozodiacal dust. However, since it is not detected by PIONIER and FLUOR observations, we consider it to be a non-detection in our statistical analysis.

\textbf{HD\,56537 ($\lambda$\,Gem)} has been observed with FLUOR before and was found to have significant K~band excess \citep{abs13}. We do not detect any H~band excess in our PIONIER data. The FLUOR accuracy on this star is slightly better ($1.7\times10^{-3}$ vs.\ $2.5\times10^{-3}$ for PIONIER), but not enough to explain the difference. \citet{abs13} discuss a problem with the diameter of this star when computed from surface brightness relations ($0.65\pm0.08\,\textrm{mas}$) compared to direct interferometric measurements ($0.835\pm0.013\,\textrm{mas}$ following \citealt{boy12} and $0.807\pm0.18\,\textrm{mas}$, new measurements in \citealt{abs13}). In the present paper, we use the diameter from the surface brightness relations in order to have a consistent diameter estimate for each target. We repeated the excess measurement for this target using the stellar diameter of $0.835\pm0.013\,\textrm{mas}$ without significant change measured in the flux ratio ($-0.51^{+0.25}_{-0.25}\%$ vs.\ $-0.42^{+0.25}_{-0.25}
\%
$). A possible explanation for the difference between the FLUOR and PIONIER measurements would be an increasing excess from H~to K~band, suggesting thermal emission and a low contribution from scattered light at both wavelengths. Another possibility would be time variability of the excess, significantly reducing the total excess flux between the two observations (October 2008 vs.\ December 2012 for the FLUOR and PIONIER observations, respectively). It would help to have another FLUOR observation confirming or ruling out the excess still being present.

\subsection{Statistics from the PIONIER sample}
\label{sect_stat}

\subsubsection{Overall detection rate}
\label{sect_over_all_rate}

\begin{figure}
 \centering
 \includegraphics[angle=0,width=\linewidth]{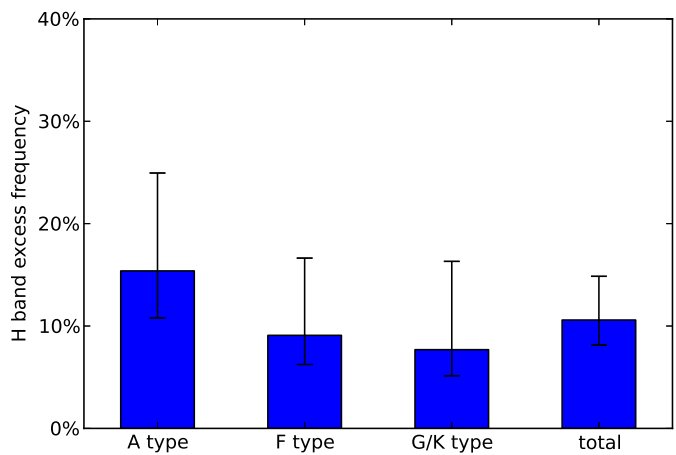}
 \caption{Detection rate of exozodiacal dust for stars of different spectral types.}
 \label{fig_hist_spt}
\end{figure}

We find a total detection rate of $10.6^{+4.3}_{-2.5}\%$ of H~band excesses that can be attributed to hot exozodiacal dust. This detection rate is less than half as high as the rate found by \citet{abs13}, a fact we discuss in Sect.~\ref{sect_compare_PIONIER_FLUOR}. Our detection rate is consistent with the result found by \citet{mil11} using KIN in N~band. However, given the different wavelength, sensitivity to different dust populations, and the large statistical error bars, drawing clear conclusions from it is not possible without detailed knowledge about the systems detected.

The detection rate obtained in this survey is similar to the detection rates for cold debris disks \citep[e.g.,][]{bei06, bry06, su06, eir13}. The dust observed in these disks, however, can be readily explained by steady-state models in which it is continuously replenished by collisions between large, km-sized planetesimals. If the H~band excesses observed in this sample were produced by the collisional evolution of planetesimals to produce dust in a similar manner, the planetesimals would have to be very close to the star, at the very least within the field of view of PIONIER. Collision rates increase with decreasing orbital timescales. \citet{wya07b} show that this leads to a maximum mass -- hence fractional luminosity -- of dust that can be produced in steady state, as a function of the orbital distance and age of the system. According to these estimates, for instance, the maximum fractional luminosity of a disk at 1\,AU and an age of 100\,Myr is $1.6\times10^{-6}$. Given only one measurement of the flux 
ratio for each available target and even with the spectrally dispersed data as discussed in Sect.~\ref{sect_spectral}, only weak constraints can be put on the fractional luminosity of the detected exozodiacal dust systems with only lower limits, typically on the order of $10^{-4}$, possibly assuming thermal emission (considering scattered light, the limits would be even larger). A realistic model of the exozodiacal dust around Vega, which is representative of the detections in this work, has been first produced by \citet{abs06} who find a fractional luminosity of $5\times10^{-4}$. These values are clearly inconsistent with the maximum levels estimated by \citet{wya07b} for any reasonable range of parameters. Thus, the excess emission observed in this survey cannot derive from dust produced locally in a steady-state collisional cascade for the ages of these stars. Alternatively, we could be observing a transient phenomena \citep{wya07b, ken13} or the aftermath of a large collision \citep[e.g.,][]{lis08,
lis09, jac14}. \citet{bon13b}; however, we show that it is unlikely that we observe the aftermath of dynamical instabilities in such a high proportion of planetary systems. The potential origin of the hot dust in an outer debris disk is discussed in Sect.~\ref{sect_rate_cold_disk}.

A potential scenario for decreasing the removal rate of dust grains from the system and, thus, to reduce the dust production rate needed to explain the presence of the dust would be the trapping of nano grains ($\sim10\,\textrm{nm}$ in size) in the magnetic fields of the host stars \citep{cze10, su13}. \citet{su13} suggest that this scenario is responsible for the hot excess seen around Vega, but the extension to A type stars is not obvious because the topology of their magnetic fields is not well known. While the dust in this scenario would still originate in an exozodiacal disk, alternative scenarios explaining the near-infrared excess, such as free-free emission produced by stellar winds, have been discussed as well \citep{abs08}. This has, however, been ruled out as an explanation for the near-infrared excess around Fomalhaut by \citet{men13}.

\subsubsection{Detection rate vs.\ spectral type}

\begin{figure}
 \centering
 \includegraphics[angle=0,width=\linewidth]{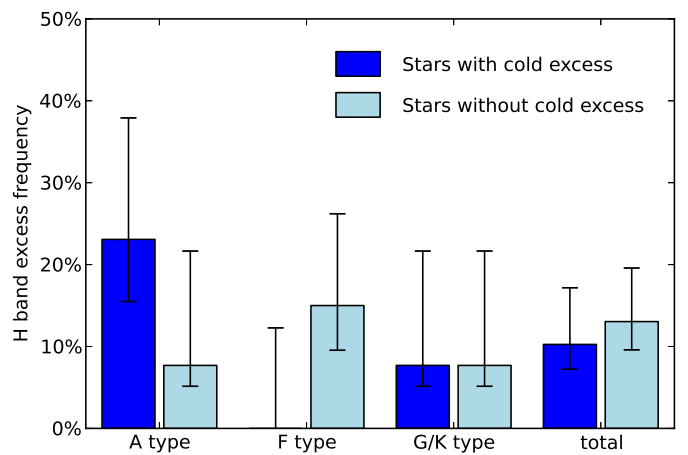}
 \caption{Detection rate of exozodiacal dust vs.\ the presence of a detected debris disk.}
 \label{fig_hist_dd}
\end{figure}

Figure~\ref{fig_hist_spt} shows the detection rate of exozodiacal dust for the different spectral type bins considered. The detection rate is decreasing toward late type stars, similar to the behavior of debris disks \citep{su06, bry06, gau07, eir13}. However, given the large statistical uncertainties (based on binomial probability distribution), this trend is only tentative.

\subsubsection{Detection rate vs.\ presence of a debris disk}
\label{sect_rate_cold_disk}

The correlation between stellar spectral type and detection rate of exozodiacal dust is similar to that of debris disks. This raises the question whether the origin of the hot and cold dust is the same population of colliding planetesimals, some of which have been transported closer to the star. This hypothesis can be tested by searching for correlations between the presence of exozodiacal dust and of a debris disk.

Given the statistical uncertainties ,there is no significant correlation between the incidence of hot and cold dust (Fig.~\ref{fig_hist_dd}). This would suggest that the two phenomena do not have a common origin. However, we are only able to detect the brightest exozodiacal dust systems and the most luminous debris disks (e.g., with \textit{Herschel} $\sim$10 times more luminous than our Kuiper belt, \citealt{vit10}). Furthermore, potentially important mechanisms, such as dust trapping in a planetary system \citep[e.g.,][]{sta08} or by stellar magnetic fields \citep{su13} or realistic treatment of sublimating dust particles \citep{leb13, vanlie14}, have not been considered in the theoretical investigation. Thus, a faint, undetected debris disk might be able to produce significant amounts of exozodiacal dust. Migration of a planet into an outer, faint belt has also been shown to potentially produce enough hot dust to be detectable by our observations and would not require a correlation between observable hot 
and cold dust \citep{bon14, ray14}.

\subsubsection{Detection rate vs.\ stellar age}
\label{sect_stat_age}

Another well known correlation for debris disks is the decrease in dust mass with stellar age, which translates into a drop in the detection rate and disk brightness with increasing age \citep{rie05, su06}. This has been attributed to the continuous mass loss due to the collisional evolution of the disk \citep[e.g.,][]{wya07a, loe08}. We already concluded that the high levels of hot dust found are unlikely to be consistent with steady-state collision evolution. Checking for a similar correlation between the detection rate and brightness of exozodiacal systems with stellar age, we can test this hypothesis further.

\begin{figure}
 \centering
 \includegraphics[angle=0,width=\linewidth]{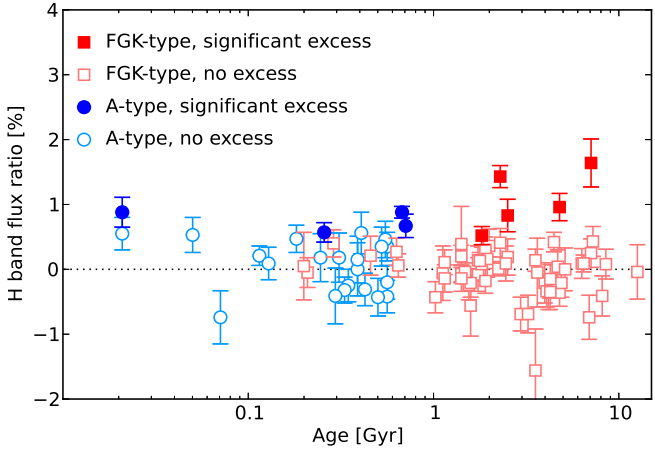}
 \caption{Excess due to exozodiacal dust as a function of stellar age in the PIONIER sample.}
 \label{fig_excess_age}
\end{figure}

\begin{figure}
 \centering
 \includegraphics[angle=0,width=\linewidth]{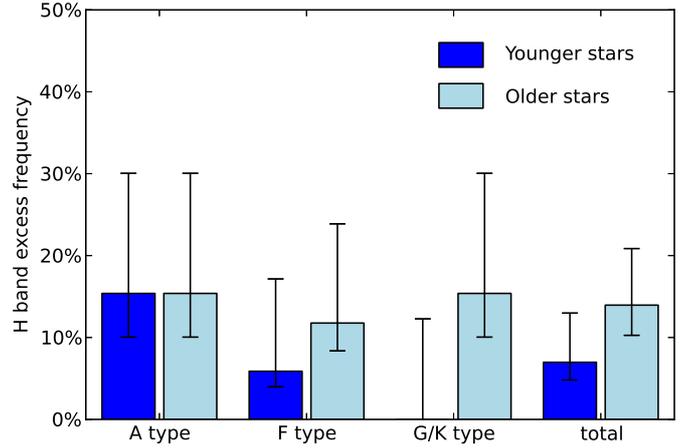}
 \caption{Detection rate of exozodiacal dust for stars younger and older than the median age of each spectral type bin. For the total bin, the age bins derived from the individual spectral type bins are considered, not a median age of the whole sample.}
 \label{fig_fraction_age}
\end{figure}

We make this test in two different ways. First, we plot the flux ratio as a function of the stellar age. We divide our sample into early-type stars (spectral type A) and stars of later spectral types, accounting for differences in the stellar properties, such as stellar luminosity, winds, and magnetic fields potentially affecting the dust evolution. Second, we investigate the excess detection rate with respect to the stellar age. For each spectral type bin (A, F, or G/K), we divide the sample into stars younger than and stars older than the median age in the bin, 0.34\,Gyr, 1.95\,Gyr, and 4.47\,Gyr for A, F, and G/K type stars, respectively. The detection rate in the young and the old star samples are compared. Finally, we perform the same analysis on all stars together, but keeping the old and young categories of the stars as indicated by the median ages in their \emph{respective} spectral type bins. We do not perform this analysis for all stars together, with \emph{one} median age for all stars. This would 
put all A-type stars in the young bin and most of the late type stars in the old bin. While such an analysis would be useful if the dust evolution depends on time in general, not on the stellar evolution, it would be heavily biased in the present case by the higher detection rate around A-type stars.

There is no clear correlation of the excesses with the age of the systems visible in Fig.~\ref{fig_excess_age}. However, there is a tentative correlation between the detection rate and the age for F and G/K type stars (Fig.~\ref{fig_fraction_age}). Considering the F and GK spectral type bins together, a $\chi^2$ test yields a probability of $\sim0.75$ that there is a real correlation between the stellar age and the excess detection rate. While a similar correlation is not visible for A-type stars, the detection rate seems to increase with age for stars of later spectral type. This is the opposite of what would be expected in the case of steady-state evolution, but does not necessarily contradict planet-induced instabilities. A similar trend is visible in the combined sample (keeping the age bins derived for the individual spectral type bins) with a probability of a real correlation with age of $\sim0.75$. Because in this case the age bins are linked to 
the stellar main sequence life time rather than absolute 
time, this might suggest that the circumstellar excess emission is caused by a stellar phenomenon rather than a circumstellar one. We note, however, that such a phenomenon would have to be very similar over a wide range of stellar spectral types and that no such phenomenon is known. A different explanation could be that the time scale of the circumstellar phenomenon (e.g., the dust evolution) that leads to the excess at older ages depends on the properties of the star. This would be the case, for example, for the Poynting-Robertson time scale, which decreases for a given dust species with increasing stellar luminosity and mass.

\subsubsection{Correlation with presence of planets}

In our sample there are 14 stars for which the detection of an exoplanetary system has been reported. All planets are located within a few AU from their host stars, near the region where the dust must be located for our exozodi detections. In our sample we find no correlation between the presence of exozodiacal dust and of planets in the system.

\subsubsection{Correlation with stellar rotation}

\citet{abs13} suggest that part of the high detection rate of hot circumstellar emission for A-type stars might be explained in part by outflows due to rapid rotation. To investigate this scenario, one can search for a correlation between high rotational velocities and a high disk-to-star contrast. We follow for the A-type stars in our sample the same approach as \citet{abs13}. We find no correlation and note in particular that there is a large number of rapid rotators without detected excess.

\subsection{Analysis of spectrally dispersed data}
\label{sect_spectral}

While our analysis so far followed closely the approach used by \citet{abs13} due to the very similar kind of data available, the \textit{SMALL} spectral dispersion of our PIONIER data with three spectral channels across the H~band allows us to also investigate the spectral slope of the excess for detected exozodiacal systems. Since the spectral channels are considered to be correlated in our contrast measurements, considering all together does not reduce the uncertainties on the combined excess measurement and, vice versa, the uncertainties in the single channels are not significantly larger than for the combined measurement. Therefore, for significant excess detections, the spectrally dispersed data may allow for putting constraints on the H~band colors of the excesses and, thus, of the location and nature of the emission.

\subsubsection{Approach}

\begin{figure}
 \centering
 \includegraphics[angle=0,width=\linewidth]{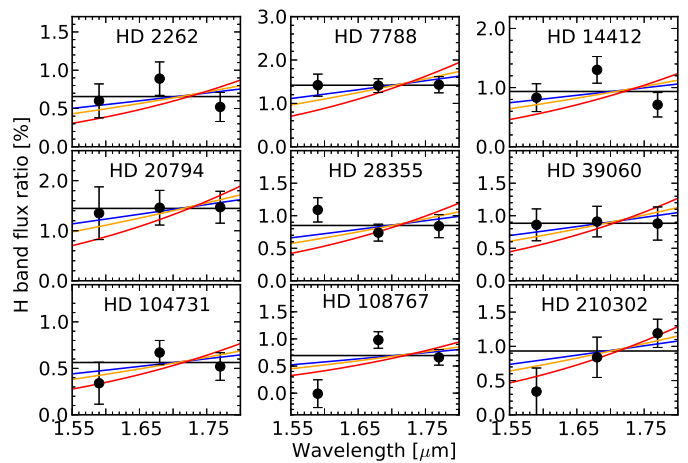}
 \caption{Disk-to-star contrast as a function of wavelength for the detected excess sources. The \emph{black} line illustrates the case of constant contrast, while the blue, orange, and red lines illustrate the cases of pure black body emission with temperatures of 2000\,K, 1500\,K, and 1000\,K, respectively.}
 \label{fig_colors}
\end{figure}

The flux ratio vs.\ spectral channel for the nine sources with detected excess is shown in Fig.~\ref{fig_colors} (first nine rows). We fit curves for thermal black body emission for four temperatures, 500\,K, 1000\,K, 1500\,K, and 2000\,K, to the data. The error-weighted average of the three spectral channels represents the case where the circumstellar emission detected follows the spectrum of the host star (i.e., constant contrast with wavelength). As can be seen in Fig.~\ref{fig_colors}, the contrast is fairly constant over the three spectral channels for most targets. This would be the case for pure gray scattered light emission from dust grains, suggesting that this makes a significant contribution to the total emission. Only the last two targets exhibit a tentative slope that might suggest thermal emission. However, for all targets we are able to rule out neither pure black body emission nor pure scattered light emission based on this qualitative analysis.

\begin{figure}
 \centering
 \includegraphics[angle=0,width=\linewidth]{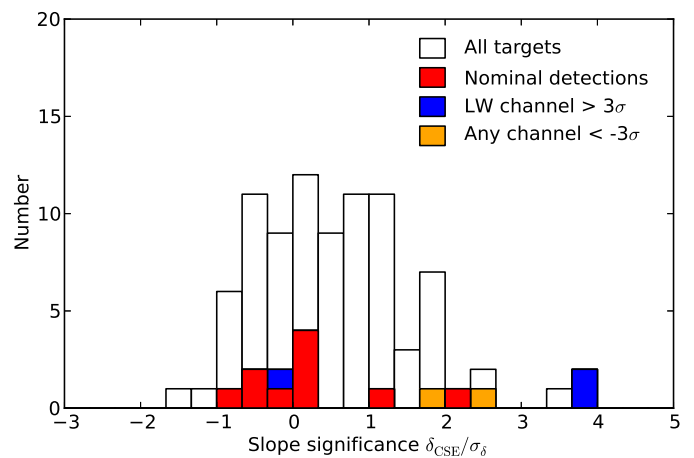}
 \caption{Distribution of the spectral slopes of the flux ratios. The white bars represent the whole sample, red bars represent nominal, broad band detections, blue bars represent targets with excess only in the long wavelength spectral channel, and orange bars represent stars with negative ($<-3\sigma$) excess in one or more channels. White bars can be hidden by colored bars, while colored bars are stacked; i.e., they cannot hide each other. The three targets with significantly positive slopes are HD\,172555, HD\,182572, and HD\,210049. The star with $a/\sigma_a \sim 0$ but with excess only in the long wavelength channel is HD\,45184. These cases are discussed in Sect.~\ref{sect_results_spectral}.}
 \label{fig_hist_slopes}
\end{figure}

To carry out a more systematic analysis of the spectral behavior of the contrast for all our targets -- be it with or without significant broad band excess -- we proceed in two ways. First, we fit a straight line $f_\textrm{CSE}(\lambda) = a\lambda + f_0$ to the three spectral channels in order to derive the spectral slope $a$ of the contrast and the corresponding uncertainty $\sigma_a$. We prefer this over a black body fit to the data because we suspect that scattered light makes a significant contribution to the emission, as discussed above. In this case, any black body temperature derived would be meaningless. In contrast, the spectral slope $a$ is a purely empirical quantity that does not need any assumptions on the underlying physics but is handy for quantitatively investigating the significance of the spectral behavior. The spectral channels are correlated, and systematic uncertainties, such as 
the uncertain diameter of the 
star or of the calibrator, affect them in the same way. These systematic uncertainties are therefore not considered in the present fit. We validate this by plotting the histogram of $a/\sigma_a$, i.e., the significance of the slopes for all targets (Fig.~\ref{fig_hist_slopes}). The histogram is well behaved with a standard deviation of 0.87, not including the three stars with significantly positive slopes. There is a small offset toward positive slopes, which we rate as insignificant. We also compute the spectral slope in the H~band of the black body curves fitted to the data of the excess stars in order to evaluate whether any black body temperature can clearly be ruled out by the spectral slope measured (Table~\ref{tab_slopes}). The slope $a$ of the flux ratio depends not only on the black body temperature, but also on the flux ratio itself, as well as on the spectral slope of the stellar photosphere emission. Thus, for a given blackbody temperature, the spectral slope of the flux ratio is expected 
to be different for different targets and excesses. For gray scattered stellar light, i.e., where the spectrum of the excess is the same as that of the star, the spectral slope of the flux ratio is $a=0$. A $3\sigma$ criterion is used to check whether the measured slope is consistent within the uncertainties with the predicted slope for different temperatures or a zero slope for scattered light emission.

Second, we search for significant excesses in only one or two spectral channels. Therefore, we consider the full uncertainties, including systematic calibration uncertainties. Excesses in the long-wavelength channel(s) might be expected if the excess is just starting to be significant in the middle of the H~band. In this case, we would expect the targets to also exhibit some positive slope of the measured flux ratio.

\subsubsection{Discussion}
\label{sect_results_spectral}

\begin{table}
\caption{Spectral slopes of the detected excesses}
\label{tab_slopes}      
\centering          
\begin{tabular}{ccccccc} 
\toprule                                                     
HD & $a$ & $\sigma_a$ & $a_\textrm{2000\,K}$ & $a_\textrm{1500\,K}$ & $a_\textrm{1000\,K}$ & $a_\textrm{500\,K}$ \\
\midrule
2262   & -0.63 & 1.49 & 1.01 & 1.48 & 2.28  & 3.46  \\
\textbf{7788}   & 0.04  & 1.67 & 2.06 & 3.11 & 5.00  & (8.46)  \\
\textbf{14412}  & -0.93 & 1.69 & 1.27 & 1.95 & 3.14  & (4.94)  \\
\textbf{20794}  & 0.50  & 2.41 & 1.94 & 2.98 & 4.80  & (7.99)  \\
\textbf{28355}  & -1.13 & 1.41 & 1.33 & 1.96 & (3.12)  & (5.25)  \\
39060  & 0.11  & 1.89 & 1.41 & 2.09 & 3.32  & 5.42  \\
104731 & 0.39  & 1.38 & 0.82 & 1.23 & 1.97  & 3.30  \\
108767 & 1.47  & 1.35 & 1.14 & 1.63 & 2.50  & 4.00  \\
210302 & 4.61  & 2.15 & 1.37 & 2.06 & 3.32  & 5.63  \\
\midrule
\textbf{172555} & \textbf{8.02}  & 2.12 & (1.37) & 2.12 & 3.65  & 7.25  \\
\textbf{182572} & \textbf{3.97}  & 1.06 & (0.29) & (0.48) & 0.87  & 1.81  \\
\textbf{210049} & \textbf{10.50} & 3.10 & (0.58) & (0.94) & 1.74  & 3.81  \\
\bottomrule
\end{tabular}
\tablefoot{Slopes and uncertainties are listed in $\%/\mu\textrm{m}$. In addition to the measured values, we list for all targets the spectral slopes of purely thermal excesses with black body temperatures of 2000\,K, 1500\,K, 1000\,K, and 500\,K. For targets marked in bold face, constraints on the emission mechanism or the black body temperature are possible. Measured slopes marked in bold face are significantly different from a zero slope, i.e., with scattered light emission, while black body slopes in parentheses can be ruled out. The three last targets are the tentative detections of circumstellar excess based on the spectral analysis, but are not detected as broad band excesses by our original approach.}
\end{table}

The results of this analysis are shown in Figs.~\ref{fig_colors} and~\ref{fig_hist_slopes}, and Table~\ref{tab_slopes}. It is obvious that -- within the $3\sigma$ uncertainties -- the slopes for all detected broad band excesses are consistent with a constant contrast with wavelength and most of the black body temperatures considered. Neither black body emission nor pure scattered light can be ruled out. However, most targets are better fit by a constant contrast than by a black body of realistic temperature. Only two sources, HD\,108767 and HD\,210302, are better fit by a black body. A clear conclusion would require data of higher precision or over a wider wavelength range. If real, the constant slope would be indicative of very hot dust or of the emission when dominated by scattering rather than thermal emission. Both scenarios have been investigated by \cite{def12} using the example of $\beta$\,Pic (HD\,39060), which exhibits similar behavior. In particular, the possibility has been 
discussed that the near-infrared excess emission is produced by forward scattering in the 
outer debris disk seen edge-on. Modeling this scenario constrained its contribution to the total near-infrared excess to be less than $\sim$50\% for $\beta$\,Pic. The debris disks around the other stars for which we detect near-infrared excesses are significantly fainter than the one around $\beta$\,Pic. In six of nine cases, a debris disk has not even been detected in the far infrared. Thus, we conclude that this scenario cannot be responsible for the excesses detected. Concluding whether the emission is dominated by thermal emission or scattered light is not possible based on the available data on our detected excesses.

\begin{figure}
 \centering
 \includegraphics[angle=0,width=\linewidth]{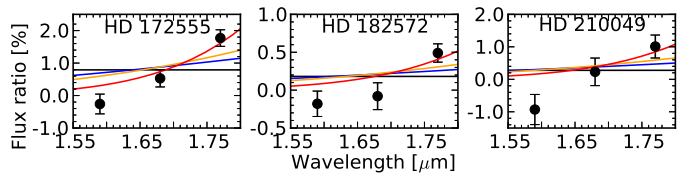}
 \caption{Same as Fig.~\ref{fig_colors} but for the three tentative detections based on the spectral slope and excess in the long wavelength channel only. The blue, orange, and red lines illustrate the cases of pure black body emission with temperatures of 1500\,K, 1000\,K, and 500\,K, respectively.}
 \label{fig_colors_more}
\end{figure}

There are, however, four more targets found not to have significant broad band emission but to exhibit a significant slope of the contrast or a significant excess in the long wavelength channel only (or both). Two of these sources, HD\,172555 and HD\,182572, fulfill both criteria, so we consider them as tentative excess detections. The tentative detection of HD\,172555 is particularly interesting thanks to the strong, warm excess this star is known to exhibit in the mid infrared (e.g., \citealt{lis09}). HD\,210049 has a significant slope, but we measure only a contrast of $2.9\sigma$ on the long wavelength channel. Since this contrast is nearly at a significant level and the slope is significant, we consider this target as a tentative detection as well. Furthermore, the contrast in the short-wavelength channel is $(-0.93\pm0.46)\%$, which together with the significant slope, suggests that in this case the absolute calibration error results in an underestimation of the contrast in all three channels. The 
source with a significant excess in the long wavelength channel but with no significant slope is HD\,45184. Here, the slope is rather constant, and we measure a contrast above $2\sigma$ on each of the three spectral channels. The reason the contrast in the last spectral channel appears significant is that the uncertainty is at its lowest here, not that the contrast is at its highest. Thus, and given that we also have two targets with negative contrast in single channels (which cannot be real), we conclude that the detection for this target is insignificant. The spectrally dispersed contrast for HD\,172555, HD\,182572, and HD\,210049 is shown in Fig.~\ref{fig_colors_more}. Their spectral slopes are listed in Table~\ref{tab_slopes}. For these targets we can rule out both purely scattered light emission, as well as very hot black body thermal emission, based on the spectral slopes. Clearly, these targets deserve follow-up observations, either deeper in H~band or at longer wavelength. Also, a deep observation 
with 
the \textit{LARGE}, seven-channel spectral resolution of PIONIER would help to put stronger constraints on the potential presence of an excess around these stars.

\begin{figure}
 \centering
 \includegraphics[angle=0,width=\linewidth]{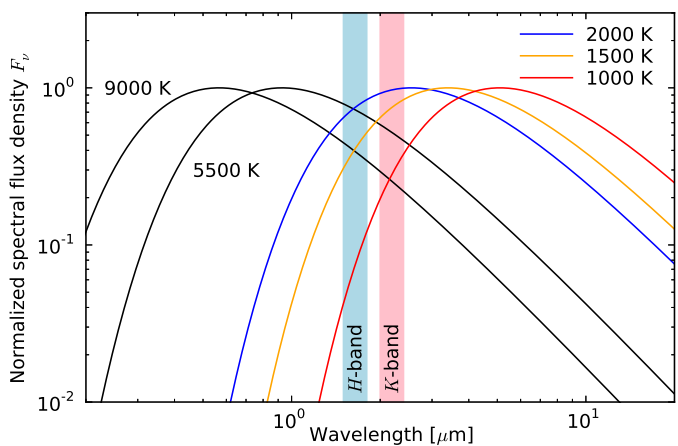}
 \caption{Spectral behavior of black body emission of different temperature. The black lines show black body curves for typical stellar temperatures in our sample, while colored lines show curves for typical sublimation temperatures of dust \citep[e.g.,][]{leb13, vanlie14}.}
 \label{fig_blackbodies}
\end{figure}

\section{Merging the \textit{VLTI}/PIONIER and \textit{CHARA}/FLUOR samples}
\label{sect_comp_chara}

\subsection{Comparison}
\label{sect_compare_PIONIER_FLUOR}

One of the goals of the PIONIER observations was to extend the sample of stars searched for exozodiacal dust by \textit{CHARA}/FLUOR toward the south by merging the two samples, in order to increase the number of targets observed and of excesses detected and thus to improve statistics. Therefore, it is mandatory to compare the two samples and to check whether they are compatible. While target selection and detection strategy for the two samples are very similar, the main difference is in the observing wavelength. PIONIER operates in H~band ($1.65\um$) and FLUOR in K~band ($2.2\um$). 

\begin{figure}
 \centering
 \includegraphics[angle=0,width=\linewidth]{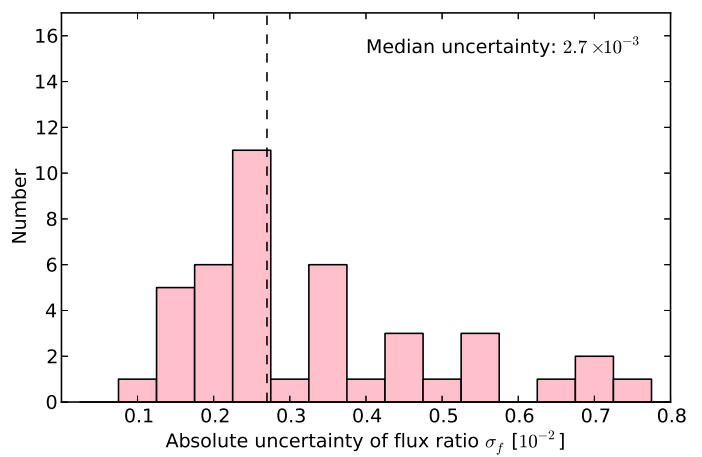}
 \caption{Sensitivity distribution of the \textit{CHARA}/FLUOR sample. For details on these data, see \citet{abs13}.}
 \label{fig_sensitivity_fluor}
\end{figure}

Figure~\ref{fig_blackbodies} shows the spectral behavior of black body emission at different temperatures. Even the emission of the hottest dust that can be present around a star is longward of the H~band. As a consequence, the flux ratio will increase toward longer wavelengths, from PIONIER to FLUOR, assuming pure blackbody thermal emission. Therefore, at a similar sensitivity to the flux ratio, the PIONIER sensitivity to circumstellar dust is lower than for FLUOR. A plot of the sensitivity distribution of the FLUOR sample is shown in Fig.~\ref{fig_sensitivity_fluor}. The median $1\sigma$ sensitivity to the flux ratio is $2.7\times10^{-3}$ compared to $2.5\times10^{-3}$ for PIONIER.

Indeed, the detection rate of PIONIER is significantly lower than for FLUOR ($10.6^{+4.3}_{-2.5}\%$ vs.\ $28^{+8}_{-6}\%$). On the other hand, the spectrally resolved data of our PIONIER detections tentatively suggest that for most of our targets, the flux ratio does not increase significantly toward longer wavelengths. A possible conclusion that needs to be confirmed by multi-wavelength observations of the same targets would be that the emission is dominated by scattered light in the H~band and that between H~and K~bands the increasing thermal emission takes over.

Another indicator as to whether the two samples are compatible would be whether observations with the two instruments arrive at consistent results for the same stars. There are three stars included in both samples, HD\,56537, HD\,69830, and HD\,71155. Only for HD\,56537 (lam Gem) has an excess been detected with FLUOR. All three targets do not show any significant excess in H~band. The non-detection for HD\,56537 has already been discussed in Sect.~\ref{sect_specific}.

\begin{figure*}
 \centering
 \includegraphics[angle=0,width=\linewidth]{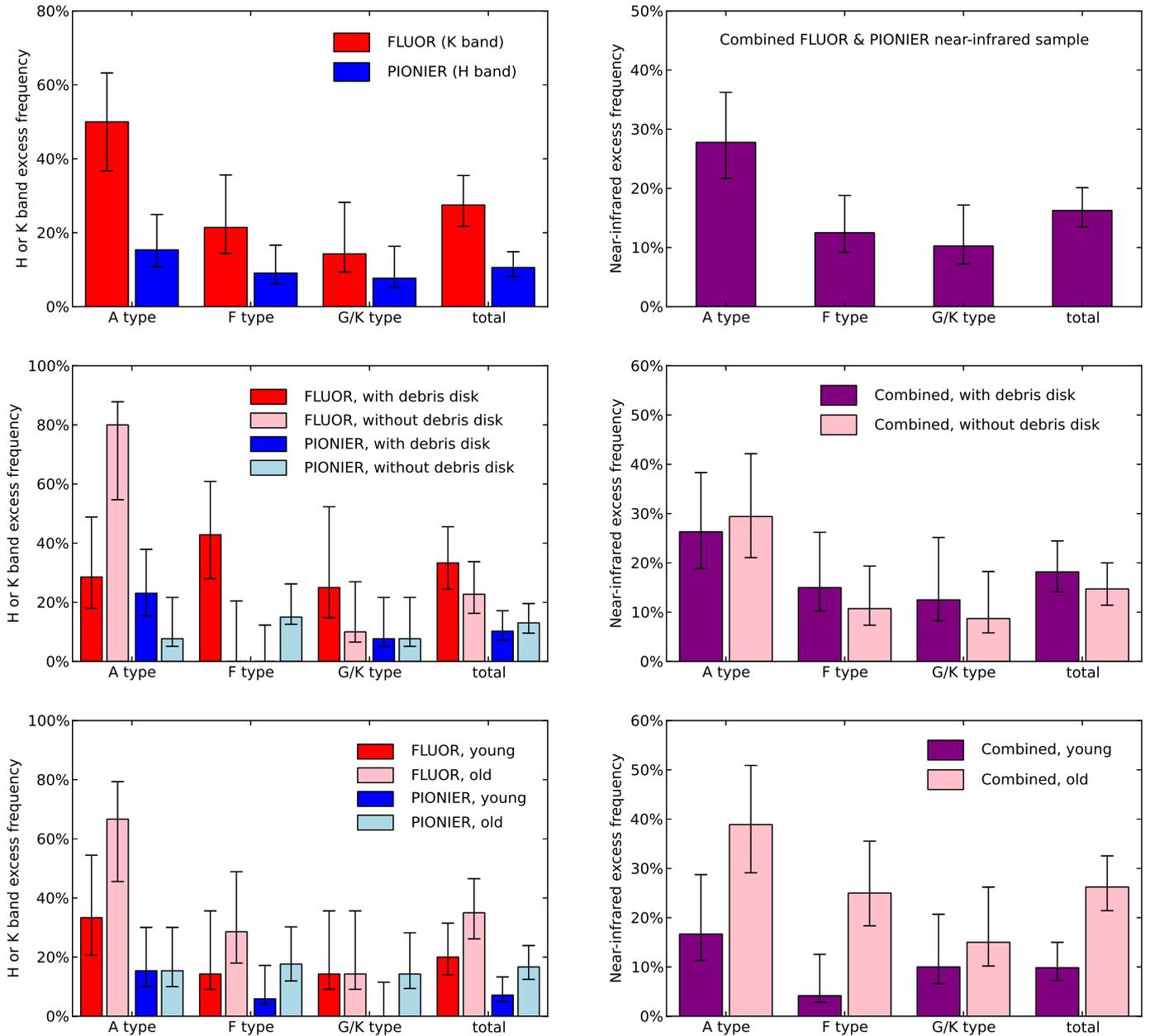}
 \caption{Comparison of the two samples and statistics performed on the merged sample.}
 \label{fig_merged_statistics}
\end{figure*}

Finally, we can compare the statistics derived from the PIONIER sample in Sect~\ref{sect_stat} with the results of \citet{abs13}. For this purpose we compare the statistics derived from the two surveys in Fig.~\ref{fig_merged_statistics}. We already noted that the detection rate is lower for the PIONIER sample by a factor of $\sim$$2$. When correcting for that (i.e., multiplying the PIONIER detection rates by this factor), the detection rate with respect to the spectral type is consistent between the two samples. A difference is clearly visible between the two samples in correlation between the presence of hot and cold dust. Here, the two samples suggest completely opposite correlations. However, again, correcting for the different overall detection rates, all detection rates in the two samples are consistent with each other given the statistical error bars. Thus, we do not consider these differences to be significant. Consistently with \citet{abs13}, we find no significant correlation of the 
excesses with stellar age.

We conclude that except for the lower detection rate in the PIONIER sample, the results are consistent.

\subsection{Merging the samples}

Merging the samples in a rigorous way is only possible taking into account color information of the detected excesses or at least on a reasonably large sample of detections that can be used as a proxy. This information is not available at the moment. Thus, we only merge the two samples and analyze the merged sample in a preliminary way.

We see two possibilities for merging the two samples. One would be to simply consider all detections in the near infrared irrespective of the observing band and thus readily co-add the H~and K~band samples. This would not require any corrections. However, the two times larger number of stars in the PIONIER sample means that the resulting combined sample would be biased toward H~band observations. The other possibility would be to correct for the lower detection rate in the PIONIER sample by multiplying all detection rates in one of the samples by a correction factor to match the rates of the other sample. The problem here is that this assumes that the difference in the overall detection rate is the only difference in the two samples that might not be true. As a consequence, the resulting sample cannot be considered to be representative of a pure H~or K~band sample either.

We decided to go for the first option by just adding up the two samples without any correction. A more sophisticated approach may be attempted later once the FLUOR sample has been extended to the same stellar flux limit as the PIONIER sample, and color information for a reasonably large sample of detections were obtained. Currently, the FLUOR sample is limited to stellar magnitudes down to $K=4$. Observations to extend the sample to $K=5$ are continuing. Attempts are in progress to constrain the near-infrared emission mechanism and temperature of exozodiacal dust through multi-wavelength data, taking advantage of PIONIER's K~band capabilities.

\begin{figure}
 \centering
 \includegraphics[angle=0,width=\linewidth]{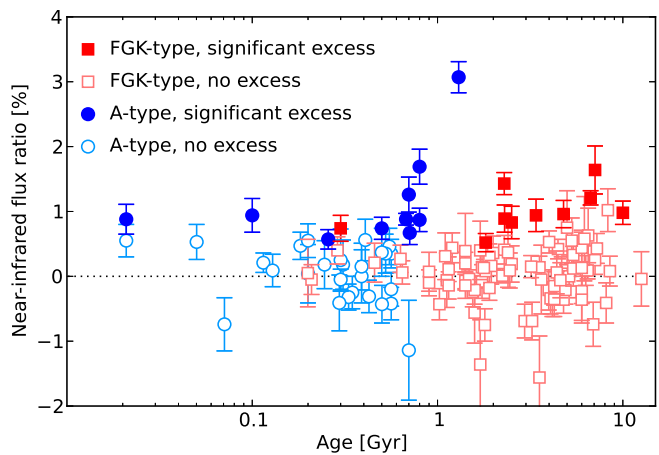}
 \caption{Excess due to exozodiacal dust as a function of stellar age for the merged sample.}
 \label{fig_merged_age}
\end{figure}

For the merged sample, we used the same statistics as presented before for the PIONIER sample (Figs.~\ref{fig_hist_spt}, \ref{fig_hist_dd}, \ref{fig_excess_age}, and \ref{fig_fraction_age}). The results are shown in Figs.~\ref{fig_merged_statistics} and~\ref{fig_merged_age}. As expected, the statistical uncertainties are reduced. With the reduced uncertainties, we widely confirm the tentative conclusions from the statistics on the PIONIER sample alone. The decreasing detection rate from early-type stars toward late type stars becomes more significant, allowing for the conclusion that this is correct. The tentative conclusion that the presence of detectable cold and hot dust in the system is not correlated becomes much more obvious in the merged sample, albeit  still with large statistical error bars. The tentative conclusion from the PIONIER data that older stars are more likely to 
harbor hot dust at a detectable level is confirmed. At least for the F-type stars, this correlation is clearly visible, and a $\chi^2$ test 
yields a probability of $0.96$ that there is a definite correlation, while this correlation is insignificant
for the A-type and GK-type spectral bins, with probabilities of $0.56$ and $0.87$, respectively. Considering stars of all spectral types together, but keeping the age bins from the individual spectral type bins, the probability that there is a definite correlation is $0.98$. The distribution of excess levels over stellar ages remains mostly unchanged. With the larger number of data points, the absence of a significant correlation becomes clearer. The tentative increase in excess levels with stellar age for A-type stars suggested by \citet{abs13} cannot be confirmed. This impression was mostly caused by one large excess at a high age (HD\,187642, alf\,Aql), while the other detected excesses show no clear correlation. This remains a single case after increasing the number of targets observed by a factor of $\sim$3 and the number of detections by a factor of $\sim$2.

\section{Summary and conclusions}
\label{sect_sum_conc}

We observed 92 nearby stars using \textit{VLTI}/PIONIER in H~band with the goal of searching for near-infrared bright circumstellar emission. This goal could be achieved for 86 main sequence stars. We reached a median sensitivity of $2.5\times10^{-3}$ ($1\sigma$) on the disk-to-star contrast for these stars. Significant extended circumstellar emission has been found around nine targets, resulting in an overall broad band detection rate of $10.6^{+4.3}_{-2.5}\%$. In addition, three tentative detection were found. The detection rate decreases with stellar spectral type from $15.4^{+9.6}_{-4.6}\%$ for A-type stars to $7.7^{+8.6}_{-2.5}\%$ for G and K type stars, similar to the known behavior of debris disks \citep{su06, bry06, gau07, eir13}. This suggests a common origin for both phenomena that may depend on the amount of solid bodies formed in planetary systems, which correlates with the stellar mass \citep{and13}. Another correlation with the mass of the host star supporting this hypothesis is that of the 
giant -planet frequency \citep{joh07} that is consistently attributed to the mass of the erstwhile protoplanetary disk by planet formation models \citep[e.g.,][]{lau04, ida05, kor06, ali11}.

Our PIONIER sample allows only tentative conclusions on correlations between the incidence of hot circumstellar emission and other properties, such as the stellar age or the presence of a debris disk, because of the limited sample size. We attempted to merge the PIONIER sample and the \textit{CHARA}/FLUOR sample first presented in \citet{abs13} in order to improve statistics. From the merged sample we find that there is no significant correlation between the presence of detectable exozodiacal dust and of a detectable debris disk. Furthermore, we find tentative evidence that the detection rate of hot exozodiacal dust increases with the age of the system. This is very surprising because any steady-state dust production mechanism from planetesimals (local or not) will remove those planetesimals from the system, thereby reducing the dust mass over time. The effect is most visible  for the F type stars in 
our sample with a median age of 1.9\,Gyr. This might indicate that the potential pile-up must occur on a Gyr 
time scale. That we do not see any significant increase in the excess levels for stars of increasing age suggests that there is a maximum amount of dust that can be trapped. We do not find any correlation of the detection rate of hot dust with the presence of known close-in planets, which seems to rule out the planetary trapping scenario, despite large statistical uncertainties. Further theoretical analysis of the proposed scenarios is necessary in order to investigate whether any of these scenarios is physically plausible.

We also analyzed the spectral behavior of the flux ratio of our targets in the H~band. For our nine broadband detections, we can exclude neither hot thermal emission of the dust nor scattering of stellar light by small dust grains as the dominant source of excess. However, eight of our detections can be explained better by scattered light (a constant slope of the flux ratio with wavelength) than by thermal emission (resulting in an increase in the flux ratio with wavelength). This might suggest that, at least for these targets, scattered light makes a significant contribution to the total emission. In addition to the broad band detected excesses, we found three more tentative detections based on their spectral slopes being significantly different from zero. We did not include those detections in the statistics because they are uncertain. However, if real, the excess of these targets is clearly dominated by thermal emission. This diversity suggests a wide diversity of architectures of exozodiacal dust systems 
in 
contrast to the dust being significantly piled up at its sublimation radius for all systems.

The impact of exozodiacal dust on future planet-finding missions has been discussed in detail by \citet{rob12} for direct imaging and by \citet{def10} for interferometric observations, both at optical and near-infrared wavelengths. \citet{rob12} find that exozodiacal dust emission at levels of our own zodiacal dust can already significantly affect the detectability of an exo-Earth for direct imaging. \citet{def10} find similar, albeit slightly weaker, constraints for an interferometric mission. The exozodiacal dust systems detected in the present survey must have a much higher fractional luminosity, typically by a factor of $\sim$1000. 

The high detection rate found already at this level suggests a significant number of fainter systems that are undetectable by our survey but still much brighter than zodiacal dust in our solar system. This might in general pose a significant obstacle for the search for exo-Earths. We can only put very limited constraints on the dust location in the detected systems. Earlier studies 
of single systems have concluded that the dust is probably very hot, close to the sublimation radius and thus not cospatial with potential exo-Earths \citep{def11, leb13}. In contrast, we find that  the H~band flux ratios of our detected systems exhibit a rather flat spectral slope, suggestive of scattered light emission. 

Although not conclusive because of large uncertainties, if real, this might suggest that the dust in most systems is colder than previously expected. At the same time, without the spectral information in our data, scattered light emission could have been misinterpreted as very hot thermal emission in previous studies, placing the dust too close to the star. Multi-wavelength information on the detected excesses, e.g., through observations in H~and K~bands and detailed modeling of these data are needed to better constrain the dominating emission process and thus the dust location. This is particularly critical since assuming an analogy to our own zodiacal dust in order to 
constrain the dust distribution is neither necessarily valid nor particularly helpful. The latter is because our knowledge about the very hot dust content of our own zodiacal disk is very limited. Only more detailed characterization of the detected systems can lead to clear conclusions on the impact of the presence of these systems on the detectability of exo-Earths.

In particular, we find no correlation between the detections of our near-infrared detected systems and of cold debris disks. In contrast, a recent analysis of KIN data on debris disk host stars suggests that the dust detected in the mid infrared is related to, but not necessarily co-located with the cold dust in debris disks (Mennesson et al., subm.). The dust detected by these observations is expected to be closer to the habitable zone. This suggests that the presence of dust in the habitable zone might be correlated more with the presence of cold dust than with the very hot and near-infrared detected dust found by our observations.

\begin{acknowledgements}
We thank P. Th\'ebault for his indispensable contribution to the EXOZODI project. S. Ertel, J.-C. Augereau, and A. Bonsor thank the French National Research Agency (ANR, contract ANR-2010 BLAN-0505-01, EXOZODI) and PNP-CNES for financial support. PIONIER is funded by the Universit\'e Joseph Fourier (UJF), the Institut de Plan\'etologie et d'Astrophysique de Grenoble (IPAG), the Agence Nationale pour la Recherche (ANR-06-BLAN-0421 and ANR-10-BLAN-0505), and the Institut National des Science de l'Univers (INSU PNP and PNPS). The integrated optics beam combiner is the result of a collaboration between IPAG and CEA-LETI based on CNES R\&T funding. The authors warmly thank everyone involved in the VLTI project. This work is based on observations made with the ESO telescopes. It made use of the Smithsonian/NASA Astrophysics Data System (ADS) and of the Centre de Donnees astronomiques de Strasbourg (CDS, A\&AS 143, 23). S. Ertel thanks K. Ertel for general support and understanding.
\end{acknowledgements}

\bibliographystyle{aa}

\bibliography{bibtex}

\begin{appendix}

\section{Chromaticism of PIONIER}
\label{app_chromat}

\begin{figure*}
 \centering
 \includegraphics[angle=0,width=\linewidth]{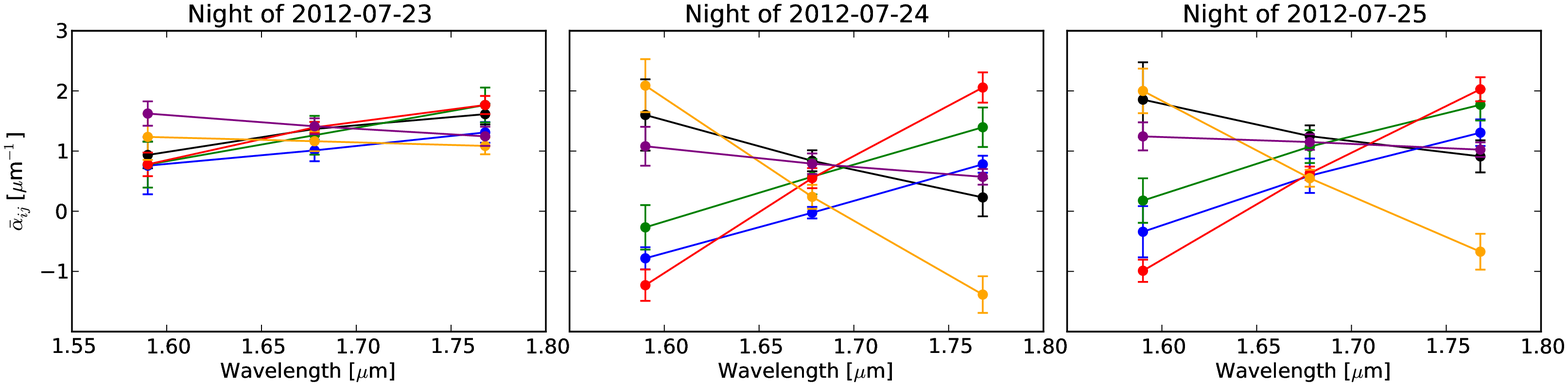}
 \caption{Spectral slopes of the transfer function of PIONIER for three illustrative nights. Line colors show the different baselines. Data points are averaged over all targets of a night, while error bars illustrate the scatter. Between the nights from July~23  and July~24 PIONIER was realigned. Between July~24 and July~25, no significant realignment was necessary.}
 \label{fig_chromaticism_1}
\end{figure*}

The chromaticism in our observations is twofold: (1) The transfer function (TF), i.e., the measured but not calibrated \sv\ of a point source given instrumental and atmospheric effects, is wavelength dependent. (2) The atmospheric transmission, the filter function, and the response of the detector are wavelength dependent, and the distribution of the flux from the three spectral channels over the three pixels of the detector depends on the alignment of the instrument. The strength of both effects may change over time and with the internal alignment of the instrument, which is usually redone before starting an observing night. If a science target and the corresponding calibrator have different spectral types, the effect of this will be a shift in the effective wavelength in each spectral channel. The resulting difference between the \sv\ measurement on a science target and the corresponding calibrator may result in systematic calibration errors that have to be characterized.

To investigate the chromaticism, we take advantage of the three spectral channel resolution data obtained during our survey for both science targets and calibrators. This provides us with \sv\ data and photometry at a low spectral resolution. The \sv\ data are corrected for the diameter of the target in order to obtain an estimate of the TF at the time of the observation. The science targets are included in this analysis to investigate the spectral type dependence of the effects studied. This is possible since we only expect a small fraction of our targets to exhibit extended emission beyond the stellar photosphere resulting only in a \sv\ drop on the order of 1\%. Thus, the whole sample can still be treated the same way as the calibrators. We fit a parabola to both the spectral shape of the TF and of the apparent flux obtained for each observation. From the data obtained in a seven-channel spectral resolution on $\beta$\,Pic \citep{def11}, we find this to be a reasonable first-order 
approximation.

From the parabola fitted to the TF data in each spectral channel $i$, baseline $j$, and for each single observation $k$, we compute a relative change of the TF with wavelength at the center $\lambda_{\textrm{c},i}$ of channel $i$:
\begin{equation}
 \label{eq_tf_slope}
 \alpha_{ijk} = \left.\frac{dV^2_{\textrm{TF},ijk}}{V^2_{\textrm{TF},ijk}~d\lambda}\right|_{\lambda = \lambda_{\textrm{c},i}}~~~.
\end{equation}
Here, $V^2_{\textrm{TF},ijk}$ denotes the TF estimated by measuring the \sv\ of a target and correcting for its diameter. Studying the dependence of $\alpha_{ijk}$ on different factors reveals that the slope of the TF depends on the baseline $j$ used (Fig.~\ref{fig_chromaticism_1}). In addition, it varies from night to night if PIONIER has been re-aligned in between. It does not change significantly if no realignment was done. Thus, we conclude that it does not significantly change during a night either. Finally, it does not depend significantly on the color of the target (spectral type). As a consequence, we can compute a median slope $\bar{\alpha}_{ij}$ 
for each night, baseline, and spectral channel by averaging all observations, respectively.

\begin{figure}
 \centering
 \includegraphics[angle=0,width=\linewidth]{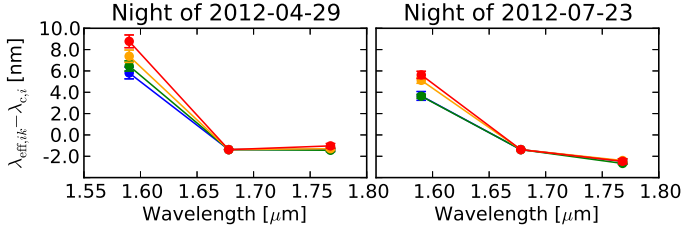}
 \caption{Difference between effective and central wavelength for two illustrative nights. Line colors show targets of different spectral type bins (blue: A type; green: F type; orange: G type; red: K type stars). Data are averaged over the targets of one spectral type and over the baselines, while the error bars illustrate the scatter. A clear trend with spectral type is visible for the first spectral channel. For the other spectral channels the trend is there as well, but barely visible given the scale of the figure.}
 \label{fig_chromaticism_2}
\end{figure}

From the parabola fitted to the spectral distribution of the apparent fluxes, we can compute an effective wavelength $\lambda_\textrm{eff,ik}$ (the barycenter of the spectral flux distribution) in each channel for each observation:
\begin{equation}
 \label{eq_lam_eff}
 \lambda_{\textrm{eff},ik} = \left(\frac{\int_{\nu_{0,i}}^{\nu_{1,i}} \nu\phi_{ik}~d\nu}{\int_{\nu_{0,i}}^{\nu_{1,i}} \phi_{ik}~d\nu}\right)^{-1}~~~,
\end{equation}
where $\nu$ is the wave number (i.e., $1/\lambda$), and $\nu_{0,i}$ and $\nu_{1,i}$ are the upper and lower boundaries of the spectral channel $i$. The quantity $\lambda_{\textrm{eff},ik}$ depends mostly on spectral type and alignment (night), but it is not expected to significantly depend on baseline or time during a night (Fig.~\ref{fig_chromaticism_2}).

Finally, we correct for the chromaticism on a per-observation, per-spectral-channel, and per-baseline basis:
\begin{equation}
 V^2_{\textrm{corr},ijk} = V^2_{ijk} \left[1-\bar{\alpha}_{ij}\left(\lambda_{\textrm{eff},ik} - {\lambda_{\textrm{c},i}}\right)\right]~.
\end{equation}
The corrections found are shown in Fig.~\ref{fig_chromaticism_2}. These corrections are applied to both calibrators and science targets, and the corrected \sv\ of the science targets are calibrated with the TF measured on the corrected \sv\ of the calibrators. The introduced correction suffers from idealization. Nonetheless, it gives a good first-order estimate of the magnitude of the chromaticism. We create two sets of calibrated data, only one of which includes the correction for chromaticism. From both data sets, we measure the excess as described in Sect.~\ref{sect_model_fitting} and compare the results. The median difference between the flux ratios measured for single targets on the data with and without applying the correction is found to be $2\times10^{-4}$. There is the expected trend from K to A type stars, which suggests that the correction works well. Besides a few cases where the correction failed owing to a bad representation of the spectral shape of the apparent flux or the TF by the parabola 
fitted 
(usually due to noisy data), the difference in the results is below $5\times10^{-4}$, clearly negligible compared to our expected accuracy of a few $1\times10^{-3}$ ($1\sigma$).

\begin{figure}
 \centering
 \includegraphics[angle=0,width=\linewidth]{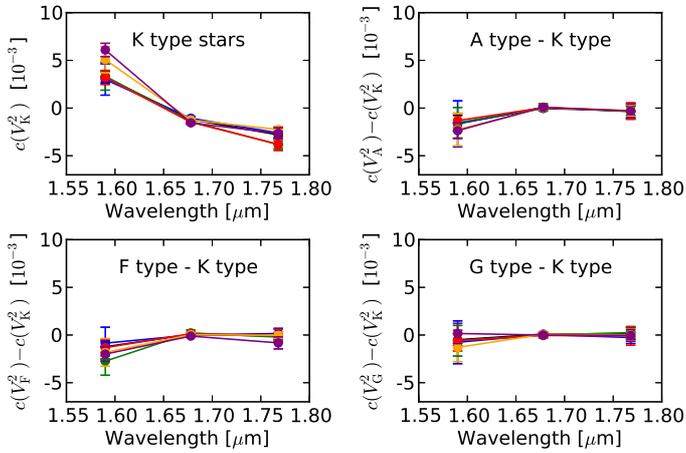}
 \caption{Absolute corrections $c(V^2)$ on the single \sv\ points derived for the chromaticism for the night of July~$23^\textrm{rd}$. We first show the correction $c(V^2_\textrm{K})$ for K type stars to illustrate the magnitude of the corrections. Then we show the difference $c(V^2_\textrm{A,F,G}) - c(V^2_\textrm{K})$ between the corrections for K type stars and stars of other spectral types, which is more illustrative of the actual error made by ignoring the correction but calibrating with K-type stars. Different lines show different baselines. The error bars illustrate the scatter of the correction for the different stars.}
 \label{fig_chromaticism_3}
\end{figure}

\end{appendix}

\begin{onecolumn}

\setcounter{table}{0}
\begin{longtable}{ccccccccccl}
 \caption{\label{tab_targets} Stellar parameters and photometry of the targets observed.}\\
 \toprule
  HD     & Sp.~T. & $d$    & $v\sin i$     & log(Age) & $\theta_{V-K}$ & $V$         & $H$           & $K$           & FIR     & FIR references  \\
  number &        & [pc]   & [km~s$^{-1}$] & [Gyr]    & [mas]          & [mag]       & [mag]         & [mag]         & excess? &                 \\
 \midrule
 \endfirsthead
 \caption{continued.}\\
 \toprule
  HD     & Sp.~T. & $d$    & $v\sin i$     & Age      & $\theta_{V-K}$ & $V$         & $H$           & $K$           & FIR?    & references      \\
 \midrule
 \endhead
 \bottomrule
 \endfoot
  142    & F7V    & $25.6$ & 11            & $ 0.44$  & $0.519^{8}$    & $5.701^{2}$ & $4.646^{76}$  & $4.474^{24}$  &  NO     & B09, E14        \\ 
  1581   & F9.5V  & $ 8.6$ & 2.3           & $ 0.60$  & $1.149^{17}$   & $4.223^{8}$ & $2.881^{20}$  & $2.815^{20}$  &  NO     & T08, E14        \\
  2262   & A5IV   & $23.5$ & 225           & $-0.15$  & $0.698^{10}$   & $3.937^{2}$ & $3.546^{20}$  & $3.523^{20}$  &  YES    & S06             \\
  3302   & F5V    & $36.2$ & 17            & $ 0.21$  & $0.493^{8}$    & $5.508^{2}$ & $4.625^{76}$  & $4.502^{26}$  &  NO     & Br06, T08       \\
  3823   & G0V    & $25.5$ & 2.3           & $ 0.83$  & $0.532^{7}$    & $5.890^{8}$ & $4.672^{16}$  & $4.486^{15}$  &  NO     & T08             \\
  4150   & A0IV   & $73.7$ & 133           & $-0.27$  & $0.457^{7}$    & $4.361^{2}$ & $4.354^{76}$  & $4.308^{27}$  &  YES    & S06             \\
  7570   & F9V    & $15.1$ & 4.3           & $ 0.55$  & $0.762^{10}$   & $4.960^{3}$ & $3.713^{20}$  & $3.666^{20}$  &  YES    & K09             \\
  7788   & F6V    & $20.4$ & 61            & $ 0.41$  & $0.739^{11}$   & $4.912^{8}$ & $3.750^{20}$  & $3.700^{20}$  &  NO     & E14             \\
  10647  & F9V    & $17.4$ & 5.5           & $ 0.26$  & $0.547^{72}$   & $5.517^{3}$ & $4.399^{234}$ & $4.340^{276}$ &  YES    & T08, E13        \\
  11171  & F3III  & $23.6$ & 58            & $ 0.02$  & $0.626^{120}$  & $4.652^{3}$ & $3.470^{180}$ & $3.890^{389}$ &  YES    & K10             \\
  14412  & G8V    & $12.7$ & 0.0           & $ 0.57$  & $0.552^{7}$    & $6.335^{4}$ & $4.694^{44}$  & $4.551^{16}$  &  NO     & T08, K09, E13   \\
  15008  & A3V    & $41.5$ & 190           & $-0.38$  & $0.541^{7}$    & $4.074^{2}$ & $3.974^{20}$  & $3.962^{20}$  &  NO     & S06, Kpc        \\
  15798  & F5V    & $25.8$ & 4.6           & $ 0.44$  & $0.770^{11}$   & $4.734^{2}$ & $3.627^{20}$  & $3.588^{20}$  &  NO     & T08             \\
  16555  & A6V    & $44.5$ & 6.6           & $-0.26$  & $0.468^{7}$    & $5.293^{3}$ & $4.600^{23}$  & $4.525^{21}$  &  NO     & P09, Kpc        \\
  17051  & F8V    & $17.2$ & 6.5           & $ 0.12$  & $0.632^{9}$    & $5.396^{2}$ & $4.130^{20}$  & $4.080^{20}$  &  NO     & Br06, T08, Kpc  \\
  17925  & K1V    & $10.4$ & 4.9           & $-0.45$  & $0.725^{10}$   & $6.044^{4}$ & $4.050^{20}$  & $4.040^{20}$  &  YES    & H08, T08, E13   \\
  19107  & A8V    & $43.1$ & 150           & $-0.74$  & $0.405^{6}$    & $5.252^{4}$ & $4.834^{42}$  & $4.741^{24}$  &  NO     & Kpc             \\
  20766  & G4V    & $12.1$ & 2.7           & $ 0.25$  & $0.676^{10}$   & $5.511^{8}$ & $4.088^{20}$  & $4.005^{20}$  &  NO     & Br06, T08, E13  \\
  20794  & G8V    & $6.02$ & 2.0           & $ 0.91$  & $1.295^{173}$  & $4.256^{8}$ & $2.709^{234}$ & $2.636^{278}$ &  YES    & K09             \\
  20807  & G0V    & $12.1$ & 2.7           & $ 0.60$  & $0.747^{11}$   & $5.228^{8}$ & $3.820^{20}$  & $3.770^{20}$  &  YES    & E13             \\
  22001  & F5V    & $21.4$ & 13            & $ 0.36$  & $0.704^{10}$   & $4.705^{4}$ & $3.740^{20}$  & $3.720^{20}$  &  NO     & Be06, Kpc       \\
  23249  & K1IV   & $ 9.0$ & 2.3           & $ 0.87$  & $1.810^{25}$   & $3.522^{3}$ & $1.539^{20}$  & $1.907^{20}$  &  NO     & Be06, K09       \\
  25457  & F6V    & $19.2$ & 18            & $-0.02$  & $0.591^{11}$   & $5.379^{3}$ & $4.342^{76}$  & $4.181^{36}$  &  YES    & H08             \\
  28355  & A7V    & $49.2$ & 105           & $-0.17$  & $0.439^{5}$    & $5.011^{2}$ & $4.570^{14}$  & $4.550^{11}$  &  YES    & S06, C08        \\
  29388  & A6V    & $45.9$ & 89            & $-0.14$  & $0.560^{7}$    & $4.262^{4}$ & $4.078^{228}$ & $3.960^{16}$  &  NO     & S06             \\
  30495  & G1.5V  & $13.3$ & 3.6           & $ 0.02$  & $0.675^{13}$   & $5.481^{3}$ & $4.116^{236}$ & $3.999^{36}$  &  YES    & T08, K09, E13   \\
  31295  & A0V    & $37.0$ & 11            & $-1.12$  & $0.448^{11}$   & $4.646^{4}$ & $4.517^{47}$  & $4.416^{47}$  &  YES    & S06             \\
  31925  & F5V    & $43.2$ & 7.2           & $ 0.35$  & $0.515^{7}$    & $5.673^{4}$ & $4.646^{76}$  & $4.479^{16}$  &  NO     & T07             \\
  33111  & A3III  & $27.2$ & 180           & $-0.41$  & $1.180^{20}$   & $2.782^{2}$ & $2.439^{204}$ & $2.380^{30}$  &  NO     & Kpc             \\
  33262  & F9V    & $11.7$ & 15            & $-0.12$  & $0.879^{98}$   & $4.708^{2}$ & $3.407^{202}$ & $3.371^{234}$ &  YES    & Br06, T08       \\
  34721  & G0V    & $24.9$ & 4.4           & $ 0.70$  & $0.494^{7}$    & $5.954^{4}$ & $4.748^{266}$ & $4.620^{20}$  &  NO     & T08, K09        \\
  38858  & G4V    & $15.6$ & 0.3           & $ 0.43$  & $0.576^{8}$    & $6.193^{7}$ & $4.499^{20}$  & $4.445^{20}$  &  YES    & Be06, K09       \\
  39060  & A6V    & $19.3$ & 13            & $-1.68$  & $0.707^{10}$   & $3.851^{2}$ & $3.500^{20}$  & $3.481^{20}$  &  YES    & S84, S06        \\
  40307  & K2.5V  & $12.8$ & 1.6           & $ 0.61$  & $0.546^{7}$    & $7.157^{7}$ & $4.968^{40}$  & $4.793^{16}$  &  YES    & E13             \\
  43162  & G5V    & $16.7$ & 5.5           & $-0.45$  & $0.496^{6}$    & $6.362^{3}$ & $4.863^{36}$  & $4.726^{16}$  &  NO     & E13             \\
  45184  & G1.5V  & $22.0$ & 2.5           & $ 0.52$  & $0.453^{6}$    & $6.366^{3}$ & $4.962^{20}$  & $4.871^{20}$  &  YES    & L09, K10        \\
  53705  & G0V    & $16.3$ & 1.6           & $ 0.93$  & $0.623^{11}$   & $5.534^{4}$ & $4.164^{20}$  & $4.140^{30}$  &  NO     & Be06, Kpc, E13  \\
  56537  & A3V    & $28.9$ & 154           & $-0.22$  & $0.651^{81}$   & $3.572^{2}$ & $3.495^{284}$ & $3.535^{262}$ &  NO     & Kpc             \\
  69830  & G8V    & $12.6$ & 1.6           & $ 0.70$  & $0.656^{9}$    & $5.945^{4}$ & $4.364^{224}$ & $4.170^{20}$  &  NO     & E13             \\
  71155  & A0V    & $38.3$ & 14            & $-0.96$  & $0.534^{7}$    & $3.881^{2}$ & $3.930^{20}$  & $3.932^{20}$  &  YES    & S06             \\
  72673  & G9V    & $12.2$ & 0.0           & $ 0.54$  & $0.597^{12}$   & $6.377^{3}$ & $4.763^{296}$ & $4.438^{36}$  &  NO     & Be06            \\
  76151  & G3V    & $17.1$ & 4.0           & $ 0.34$  & $0.537^{9}$    & $5.998^{3}$ & $4.530^{20}$  & $4.500^{30}$  &  YES    & Br06, T08, E13  \\
  76932  & G2V    & $21.3$ & 2.6           & $ 1.05$  & $0.561^{12}$   & $5.809^{3}$ & $4.389^{20}$  & $4.380^{43}$  &  NO     & Be06, Kpc       \\
  82434  & F3IV   & $18.6$ & 156           & $ 0.15$  & $1.065^{18}$   & $3.582^{2}$ & $2.700^{234}$ & $2.760^{30}$  &  NO     & C05             \\
  88955  & A2V    & $31.5$ & 10            & $-0.46$  & $0.597^{10}$   & $3.832^{2}$ & $3.758^{20}$  & $3.742^{30}$  &  YES    & Z11             \\
  90132  & A8V    & $40.5$ & 270           & $-1.15$  & $0.426^{6}$    & $5.331^{2}$ & $4.789$       & $4.686^{21}$  &  NO     & Kpc             \\
  91324  & F9V    & $21.9$ & 8.3           & $ 0.58$  & $0.793^{80}$   & $4.888^{2}$ & $3.588^{204}$ & $3.582^{214}$ &  NO     & Be06            \\
  99211  & A7V    & $25.7$ & 7.3           & $-0.25$  & $0.705^{188}$  & $4.072^{3}$ & $3.523^{570}$ & $3.546^{526}$ &  NO     & Kpc             \\
  102365 & G2V    & $ 9.2$ & 0.7           & $ 0.83$  & $0.943^{14}$   & $4.881^{8}$ & $3.377^{20}$  & $3.308^{20}$  &  NO     & Be06, Kpc       \\
  104731 & F5V    & $24.2$ & 15            & $ 0.19$  & $0.604^{8}$    & $5.153^{2}$ & $4.117^{20}$  & $4.085^{20}$  &  NO     & B06, T08        \\
  108767 & A0IV   & $27.0$ & 236           & $-0.48$  & $0.792^{11}$   & $2.953^{4}$ & $3.049^{20}$  & $3.055^{20}$  &  NO     & S06, Kpc        \\
  109787 & A2V    & $40.4$ & 296           & $-0.38$  & $0.577^{8}$    & $3.485^{2}$ & $3.715^{20}$  & $3.702^{20}$  &  NO     & Kpc             \\
  115617 & G7V    & $ 8.5$ & 3.9           & $ 0.66$  & $1.147^{129}$  & $4.727^{8}$ & $2.974^{176}$ & $2.956^{236}$ &  YES    & B06, T08        \\
  120136 & F6IV   & $15.6$ & 15            & $ 0.14$  & $0.856^{12}$   & $4.480^{2}$ & $3.400^{20}$  & $3.350^{20}$  &  NO     & T08, B09, E13   \\
  128898 & A7V    & $16.4$ & 13            & $\dots$  & $1.005^{25}$   & $3.174^{2}$ & $2.730^{30}$  & $2.740^{50}$  &  NO     & E13             \\
  129502 & F2V    & $18.7$ & 47            & $ 0.12$  & $1.027^{14}$   & $3.865^{3}$ & $2.938^{20}$  & $2.895^{20}$  &  YES    & E14             \\
  130109 & A0V    & $39.5$ & 285           & $-0.41$  & $0.613^{21}$   & $3.726^{2}$ & $3.628^{202}$ & $3.670^{70}$  &  NO     & Kpc             \\
  134083 & F5V    & $19.7$ & 44            & $ 0.17$  & $0.661^{7}$    & $4.926^{2}$ & $3.905^{20}$  & $3.880^{10}$  &  NO     & T08, Br06, E13  \\
  135379 & A3V    & $29.6$ & 68            & $-0.61$  & $0.569^{8}$    & $4.060^{2}$ & $3.890^{20}$  & $3.880^{20}$  &  YES    & M09$^\textrm{mir}$ \\
  136202 & F8IV   & $24.7$ & 4.8           & $ 0.63$  & $0.623^{85}$   & $5.048^{3}$ & $3.947^{284}$ & $4.008^{284}$ &  YES    & K10$^\textrm{mir}$ \\
  139664 & F5V    & $17.5$ & 1.8           & $ 0.02$  & $0.723^{10}$   & $4.638^{3}$ & $3.680^{20}$  & $3.660^{20}$  &  YES    & Be06            \\
  141891 & F1     & $12.3$ & 92            & $ 0.76$  & $1.433^{20}$   & $2.826^{2}$ & $2.132^{20}$  & $2.085^{20}$  &  YES    & K10$^\textrm{mir}$ \\
  149661 & K2V    & $ 9.8$ & 2.2           & $ 0.13$  & $0.776^{11}$   & $5.762^{4}$ & $3.910^{20}$  & $3.855^{20}$  &  NO     & Br06, T08       \\
  152391 & G8.5V  & $16.9$ & 3.0           & $ 0.11$  & $0.487^{8}$    & $6.649^{8}$ & $4.942^{44}$  & $4.835^{29}$  &  NO     & Br06, T08       \\
  160032 & F4V    & $21.9$ & 16            & $ 0.35$  & $0.662^{94}$   & $4.754^{4}$ & $3.702^{218}$ & $3.830^{294}$ &  YES    & E14             \\
  160915 & F5V    & $17.5$ & 12            & $ 0.42$  & $0.655^{88}$   & $4.861^{3}$ & $3.787^{270}$ & $3.877^{280}$ &  NO     & K10*            \\
  164259 & F2IV   & $23.2$ & 69            & $ 0.25$  & $0.716^{10}$   & $4.620^{4}$ & $3.700^{20}$  & $3.670^{20}$  &  NO     & Be06            \\
  165777 & A4IV   & $25.4$ & 65            & $-0.33$  & $0.717^{12}$   & $3.711^{3}$ & $3.426^{216}$ & $3.420^{30}$  &  NO     & Kpc             \\
  172555 & A7V    & $29.2$ & 116           & $-1.68$  & $0.494^{9}$    & $4.769^{2}$ & $4.251^{212}$ & $4.298^{31}$  &  YES    & R08             \\
  178253 & A2V    & $39.8$ & 20            & $-0.41$  & $0.514^{67}$   & $4.094^{2}$ & $3.915^{252}$ & $4.049^{272}$ &  YES    & M09$^\textrm{mir}$ \\
  182572 & G8IV   & $15.6$ & 2.2           & $ 0.76$  & $0.858^{18}$   & $5.152^{2}$ & $3.545^{20}$  & $3.530^{40}$  &  NO     & K10*            \\
  188228 & A0V    & $32.5$ & 89            & $-1.42$  & $0.587^{72}$   & $3.946^{2}$ & $3.762^{234}$ & $3.800^{258}$ &  YES    & S06, B13        \\
  192425 & A2V    & $47.1$ & 180           & $-0.37$  & $0.378^{5}$    & $4.940^{3}$ & $4.801^{34}$  & $4.767^{17}$  &  YES    & M09             \\
  195627 & F0V    & $27.6$ & 122           & $-0.70$  & $0.578^{65}$   & $4.749^{2}$ & $4.016^{228}$ & $4.044^{236}$ &  YES    & R08             \\
  197157 & A9IV   & $24.2$ & 150           & $-0.28$  & $0.639^{82}$   & $4.506^{3}$ & $3.692^{280}$ & $3.820^{268}$ &  NO     & P09, Kpc        \\
  197692 & F5V    & $14.7$ & 41            & $ 0.14$  & $0.949^{119}$  & $4.139^{2}$ & $3.104^{184}$ & $3.094^{262}$ &  NO     & B06, T08, Kpc   \\
  202730 & A5V    & $29.8$ & 135           & $-0.31$  & $0.518^{8}$    & $4.482^{2}$ & $4.224^{76}$  & $4.145^{26}$  &  NO     & P09             \\
  203608 & F9V    & $ 9.2$ & 3.7           & $ 0.69$  & $1.078^{15}$   & $4.223^{4}$ & $2.990^{20}$  & $2.917^{20}$  &  NO     & B06, T08, Kpc   \\
  206860 & G0V    & $18.4$ & 10            & $-0.15$  & $0.514^{10}$   & $5.953^{4}$ & $4.598^{36}$  & $4.559^{38}$  &  YES    & B06, T08, E13   \\
  207129 & G2V    & $15.6$ & 3.5           & $ 0.25$  & $0.627^{9}$    & $5.567^{2}$ & $4.200^{20}$  & $4.140^{20}$  &  YES    & T08, E13        \\
  210049 & A1.5IV & $40.0$ & 307           & $-0.51$  & $0.455^{6}$    & $4.488^{3}$ & $4.351^{20}$  & $4.349^{20}$  &  NO     & S06, Kpc        \\
  210277 & G0     & $21.3$ & 1.8           & $ 0.92$  & $0.488^{7}$    & $6.535^{4}$ & $4.957^{31}$  & $4.799^{20}$  &  YES    & E13             \\
  210302 & F6V    & $18.7$ & 14            & $ 0.35$  & $0.705^{10}$   & $4.929^{4}$ & $3.820^{20}$  & $3.780^{20}$  &  NO     & T08, E13        \\
  210418 & A1V    & $29.6$ & 144           & $-0.28$  & $0.734^{10}$   & $3.520^{2}$ & $3.380^{20}$  & $3.330^{20}$  &  NO     & S06, Kpc        \\
  213845 & F7V    & $22.7$ & 35            & $ 0.07$  & $0.523^{82}$   & $5.206^{3}$ & $4.266^{258}$ & $4.327^{326}$ &  YES    & E14             \\
  214953 & F9.5V  & $23.6$ & 4.5           & $ 0.59$  & $0.526^{7}$    & $5.991^{4}$ & $4.595^{20}$  & $4.532^{20}$  &  NO     & Kpc             \\
  215648 & F7V    & $16.3$ & 6.7           & $ 0.69$  & $1.089^{41}$   & $4.203^{4}$ & $3.078^{214}$ & $2.895^{80}$  &  NO     & Be06, K09, Kpc  \\
  215789 & A2IV   & $39.8$ & 235           & $-0.24$  & $0.899^{42}$   & $3.480^{2}$ & $3.162^{268}$ & $3.000^{100}$ &  YES    & E14             \\
  216435 & G0V    & $33.3$ & 5.7           & $ 0.65$  & $0.472^{8}$    & $6.020^{3}$ & $4.741^{20}$  & $4.711^{30}$  &  YES    & B09, K09        \\
  219482 & F6V    & $20.6$ & 7.5           & $ 0.05$  & $0.527^{7}$    & $5.649^{3}$ & $4.606^{228}$ & $4.437^{15}$  &  YES    & B06, E13        \\
  219571 & F4V    & $22.0$ & 79            & $ 0.42$  & $1.003^{14}$   & $3.992^{2}$ & $3.025^{20}$  & $2.968^{20}$  &  NO     & K10*            \\
  224392 & A1V    & $48.7$ & 20            & $-0.38$  & $0.368^{5}$    & $4.994^{2}$ & $4.949^{31}$  & $4.824^{21}$  &  NO     & D12, Z11        \\
\end{longtable}
\tablefoot{Uncertainties on the stellar diameters and magnitudes are given in $\mu\textrm{as}$ and $10^{-3}$ mag, respectively. The note ``mir'' added to some reference for the far-infrared detection means that only mid-infrared data or only upper limits in the far-infrared are available, but the detected mid-infrared excess strongly suggests the presence of excess at longer wavelengths.}
\tablebib{Distances were taken from the Hipparcos catalog \citep{per97}.\\
References for the far-infrared excesses are B09:~\citet{bry09}, B13:~\citet{Boo13}, Be06:~\citet{bei06}, Br06:~\citet{bry06}, C05:~\citet{che05}, C08:~\citet{cie08}, D12:~\citet{don12}, E13:~\citet{eir13}, E14:~this work, see Sect.~\ref{sect_sample}, H08:~\citet{hil08}, K09:~\citet{kos09}, K10:~\citet{koe10}, K10*:~\citet{koe10}, observed but no detection published, Kpc:~\textit{Herschel}/DEBRIS data, G.~Kennedy, personal communication, L09:~\citet{law09}, M09~\citet{mor09}, P09:~\citet{pla09}, R08:~\citet{reb08}, S06:~\citet{su06}, S84:~\citet{smi84}, T07:~\citet{tri07}, T08:~\citet{tri08}, Z11:~\citet{zuc11}.\\
Stellar parameters were collected from \citet{mal90, all99, ger99, fel01, ers03, mal03, gle05, val05, gra06, saf08, sou08, one09, laf10, sou10, cas11, dia11, wu11, amm12, vanbe12, zor12}.\\
Stellar ages were collected from \citet{edv93, mar95, roc98, ger99, lac99, zuc00, fel01, che01, ibu02, lam04, roc04, tho04, wri04, ben05, rie05, val05, red06, bar07, ram07, rhe07, tak07, mam08, hol09, cha10, ghe10, gon10, cas11, tet11, tre11, mal12, vic12, zor12, eir13, mal13, pac13, tsa13}.}

\end{onecolumn}

\end{document}